\documentclass[a4paper, 11pt]{article}
\pdfoutput=1  
\usepackage{jheppub} 
                     
\usepackage[T1]{fontenc}     
\usepackage{tikz, pgf} 
\usepackage{tikz-feynman,contour}
\usepackage{hyperref}  
\usepackage{amsmath,physics,float}  
\usepackage{amssymb}  
\usepackage{latexsym}
\usepackage{graphicx,xcolor}
\usepackage{subfig} 
\usepackage{empheq}
\usepackage{verbatim}
\definecolor{pink}{HTML}{FDDBC7}
\definecolor{red}{HTML}{B2182B} 
\definecolor{blue}{HTML}{4393C3}

\hbadness=\maxdimen
\vbadness=\maxdimen

\allowdisplaybreaks[1]

\title{\boldmath 
Open effective theory of scalar field in rotating plasma}
\author[1]{Bidisha Chakrabarty,}
\author[2]{Aswin P M}

\affiliation[1]{University of Southampton,
University Road,
Southampton 
SO17 1BJ,
United Kingdom}
\affiliation[2]{International Centre for Theoretical Sciences (ICTS-TIFR),
Shivakote, Hesaraghatta Hobli, Bengaluru 560089, India.}
\emailAdd{b.chakrabarty@soton.ac.uk}
\emailAdd{aswin.pm@icts.res.in} 

\abstract{We study the effective dynamics of an open scalar field interacting with a strongly-coupled two-dimensional rotating CFT plasma. The effective theory is determined by the real-time correlation functions of the thermal plasma. We employ holographic Schwinger-Keldysh path integral techniques to compute the effective theory. The quadratic effective theory computed using holography leads to the linear Langevin dynamics with rotation. The noise and dissipation terms in this equation get related by the fluctuation-dissipation relation in presence of chemical potential due to angular momentum. We further compute higher order terms in the effective theory of the open scalar field. At quartic order, we explicitly compute the coefficient functions that appear in front of various terms in the effective action in the limit when the background plasma is slowly rotating. The higher order effective theory has a description in terms of the non-linear Langevin equation with non-Gaussianity in the thermal noise.}

\begin{document}
\maketitle
\raggedbottom

\section{Introduction} 
Real time correlation functions are useful objects to study in quantum field theories.
For a field theory in vaccum or thermal state, the Lorentzian correlators can be computed by analytically continuing the Euclidean correlation functions. However for strongly coupled field theories, a direct computation of correlation functions are extremely challenging. For holographic strongly coupled conformal field theories one relies on the celebrated anti-de-sitter/conformal field theory (AdS/CFT) correspondence  \cite{Maldacena:1997re} to compute CFT correlation functions using holography. The duality in its original form was proposed to calculate Euclidean correlation functions in strongly coupled $d$ dimensional CFTs by doing computations on the weakly coupled $d+1$ dimensional AdS gravity side \cite{Gubser:1998bc, Witten:1998qj}. 
This recipe was further extended to compute Lorentzian correlation functions of the CFT using holography in \cite{Son:2002sd}. 
One of the main ingredients in this extension was to impose ingoing wave boundary condition at the horizon of AdS black holes to get the Lorentzian correlation functions of the thermal CFT. The authors of \cite{Son:2002sd} argued that there would be no boundary contribution from the horizon in the calculation of retarded bulk to boundary Green's function. The prescription was further developed for the maximally extended Kruskal spacetime by computing all bulk to boundary propagators for a two sided black hole \cite{Herzog:2002pc}. 
The corresponding CFT path integral contour corresponds to the finite temperature Schwinger-Keldysh (SK) contour with doubled degrees of freedom. 
\par
The above prescription was extensively used for the computation of real time two point functions of the CFT using holography in various scenarios \cite{Barnes:2010jp, Son:2009vu, CaronHuot:2011dr, Chesler:2011ds, Botta-Cantcheff:2018brv, Botta-Cantcheff:2019apr, de2009brownian}. However a viable prescription for computing real time $n$-point functions required a detailed description of the bulk manifold over which the path integral has to be performed to compute strongly coupled CFT correlation functions holographically. A proposal of such a bulk path integral framework was given in \cite{Skenderis:2008dh, Skenderis:2008dg}. 
Since the bulk path integral contour asymptotes to the CFT Schwinger-Keldysh contour at the boundary, the idea is to fill in the imaginary and real time segments of the CFT contour by Euclidean and Lorentzian geometries in the bulk. One has to impose appropriate matching conditions (fields and their conjugate momenta have to be continuous) at the corners of the contour where the metric changes signature. At finite temperature, the bulk manifold in this prescription has two copies of half Euclidean and half Lorentzian $AdS$ black brane geometries (glued across the initial time slice) and the two copies are then stitched together across a spacelike hypersurface. The corresponding CFT state is the thermofield double (an entangled pure state). Given the prescription of entire path integration contour on the gravity side, the generating functional of the CFT correlation functions can be obtained by integrating over the bulk fields with sources as boundary conditions where the operators inside the CFT correlator couple to the sources. 
\par
Recently Glorioso, Crossley, Liu in \cite{Glorioso:2018mmw} have given a proposal on how to handle the near horizon region of the holographic contour, that is, how to stitch the two copies of the black branes across their horizons. This is a crucial point to address since the near horizon regions are potential sources of IR divergences once interactions are turned on \cite{atmaja2014holographic}. The proposal of \cite{Glorioso:2018mmw} suggests that the bulk dual of the asymptotic Schwinger-Keldysh contour is given by a complexified double-sheeted spacetime. The two copies of the black branes in the bulk are stitched by a `horizon cap' across their future horizons. The CFT state in this case is not the thermofield double, since in the bulk the prescription involves spacetime only outside the future horizon. In \cite{Glorioso:2018mmw} the authors have computed the quadratic effective actions of probe scalar and gauge fields to test their formalism (also see \cite{deBoer:2018qqm}). 
\par
In \cite{Chakrabarty:2019aeu} the authors use the `horizon cap' boundary conditions of \cite{Glorioso:2018mmw} to calculate the quartic order effective action of a probe quark moving in a strongly coupled thermal CFT plasma (bath) using holography. This gives rise to the leading non-linear corrections to the Brownian motion of the heavy quark. If the CFT bath is sufficiently forgetful (that is when the bath correlators decay exponentially fast in time), the effective theory of the quark becomes local in time (Markovian regime). The local effective theory of the quark precisely matches the effective theory obtained in \cite{Chakrabarty:2019qcp} (see also \cite{Chakrabarty:2018dov}). 
\par
Effective theories of a quantum Brownian particle (open probe) interacting with various baths were obtained by integrating out the bath degrees of freedom \cite{Feynman:1963fq, Caldeira:1982iu, Caldeira:1982uj, Breuer:2002pc}. When the Brownian particle is linearly coupled to the bath degrees of freedom, particle's effective theory becomes quadratic. This problem has a classical stochastic description in terms of the linear Langevin equation with a Gaussian noise. In thermal equilibrium noise and dissipation terms in the Langevin equation are related by the fluctuation dissipation relation.  In \cite{Chakrabarty:2018dov} the most general form of the local effective theory upto cubic order was constructed where the authors considered a quantum Brownian particle (open system) weakly interacting with a large thermal bath (made of two sets of harmonic oscillators) via cubic interaction. This can be seen as a perturbation over the well-known Caldeira-Leggett model. The effective theory in \cite{Chakrabarty:2018dov} is described by a non-linear Langevin equation with a non-gaussian noise. If the bath has microscopic time-reversal invariance then the non-Gaussianity in the thermal noise gets related to the thermal jitter in the damping constant of the Brownian particle. This is the generalisation of the fluctuation dissipation relation. The path integral in the stochastic problem is related to an underlying Schwinger-Keldysh quantum path integral. In \cite{Chakrabarty:2019qcp}
the authors extended their work by constructing an effective theory of the particle upto quartic order.
\par
In \cite{Chakrabarty:2019aeu} the validity of the effective theory is tested for a strongly coupled bath using holography. Based on these analysis, in \cite{Jana:2020vyx} the authors study open quantum field theories using holographic Schwinger-Keldysh path integral.
The authors consider an interacting scalar quantum field theory (the system) coupled to a holographic field
theory in $d$ spacetime dimensions at finite temperature (the environment). They study the effects of integrating out
the environment and obtain the effective dynamics of the resulting
open quantum field theory in a derivative expansion in small frequency and momentum. For discussions on open quantum field theories and their features see \cite{Avinash:2017asn, Avinash:2019qga}. In \cite{1830227} the authors study fermionic open effective field theories coupled to holographic baths. They show how the holographic Schwinger-Keldysh path integral framework leads to boundary correlators that automatically satisfy fermionic Kubo, Martin, Schwinger (KMS) relations \cite{Kubo:1957mj, Kubo_1966}. Computing the bulk on-shell action they obtain the influence phase (the effect of integrating out the bath degrees of freedom on the probe action) of the probe fermion in a derivative expansion. In \cite{1830471} the authors extend this analysis to charged fermions probing Reissner-Nordstr\"om black hole background. 
\par
In this paper we extend the analysis of \cite{Jana:2020vyx} to rotating BTZ black hole. We will consider an interacting probe scalar quantum field theory (the system) coupled to a strongly coupled two dimensional rotating CFT plasma (bath) at finite temperature and chemical potential.\footnote{In \cite{Atmaja:2012jg, atmaja2014holographic} the author studied holographic Brownian motion (linear Langevin dynamics) of a heavy quark in a two dimensional rotating plasma.} In the bulk the probe scalar field is coupled to a rotating BTZ black hole \cite{Banados:1992wn, carlip19952+}. We consider a natural extension of the holographic Schwinger-Keldysh path integral framework of \cite{Glorioso:2018mmw} to the rotating BTZ black hole with two horizons. We study the effects coming from integrating out the bath degrees of freedom on the effective action of the open scalar field theory. 
The open scalar field gets dragged due to the rotation of the background thermal plasma and is described by a generalised nonlinear Langevin equation. 
\par
Following is the outline of the paper. We briefly review the gravitational Schwinger-Keldysh path integral framework of \cite{Glorioso:2018mmw} and its natural extension to the rotating BTZ black hole in section \ref{sec:Holographic SK}. In section \ref{sec:Probe scalar} we study the dynamics of a massive probe scalar field $\Phi$ in BTZ background. We solve the scalar wave equation with appropriate boundary conditions to find the ingoing and outgoing bulk to boundary green's functions. We compute the influence functionals (quadratic and higher order) of the open scalar field holographically in section \ref{sec:Influence functional}.\footnote{In \cite{Son:2002sd}, the authors computed quadratic effective theory and two point functions of the scalar field.} In presence of rotation, we study the stochastic dynamics governed by a non-linear Langevin equation resulting from the open effective field theory in section \ref{sec:Nonlinear Langevin dynamics}. We also find the generalised fluctuation dissipation relations obeyed by the parameters of the non-linear Langevin equation. Finally we summarise our results and discuss possible future directions in section \ref{sec:Conclusion}.
\par
Various technical details of the computation are relegated to the appendices. In appendix \ref{sec:integrals1} we provide details of the computation of quadratic influence functional of the open scalar field in `retarded-advanced' (RA) basis. In appendix \ref{sec:derivative expansion} we discuss the massive scalar field solution in a derivative expansion in small frequency and momentum. The derivative expansion of the solution is particularly useful in computing higher order effective theory of the open scalar field. 
\section {Holographic SK prescription}
\label{sec:Holographic SK}
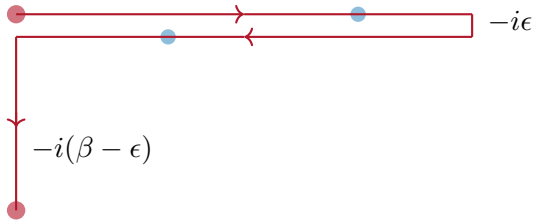
\begin{figure}[ht]
\begin{center}
\begin{tikzpicture}
\begin{scope}[thick]
\filldraw[color=red!60, fill=red!60, very thick](-0.0,-2.6) circle (.1);
\filldraw[color=red!60, fill=red!60, very thick](-0.0,-0.0) circle (.1);
\filldraw[color=blue!60, fill=blue!60, very thick](4.5,0.0) circle (.08);
\filldraw[color=blue!60, fill=blue!60, very thick](2.0,-0.3) circle (.08);
\draw [red,->](0,0.0) --(3.0,0.0);
\draw [red](3.0,0.0) --(6.0,0.0);
\draw [red](6.0,0.0) --(6.0,-0.3);
\draw [red,->](6.0,-0.3) --(3.0,-0.3);
\draw [red](3.0,-0.3) --(0.0,-0.3);
\draw [red,->](0,-0.3) --(-0.,-1.5);
\draw [red](0,-1.3) --(-0.,-2.6);
\end{scope}
\node at (6.5,-0.1) {$-i\epsilon$};
\node at (1.,-1.8) {$-i(\beta-\epsilon)$};
\end{tikzpicture}
\end{center}
\caption{Schwinger-Keldysh imaginary time contour with an initial thermal state.}
\label{CFT contour}
\end{figure}

Before we start reviewing the holographic SK path integral framework of \cite{Glorioso:2018mmw}, let us describe the path integral contour in the boundary CFT on which the CFT correlation functions are defined. For the system prepared in an initial thermal state, the Schwinger-Keldysh time contour as shown in Fig \ref{CFT contour} has imaginary (vertical leg) as well as real legs (horizontal lines). The initial and final points of the contour have to be identified on a thermal circle with periodicity $\beta$, where $\beta$ is the inverse temperature of the CFT.
\par
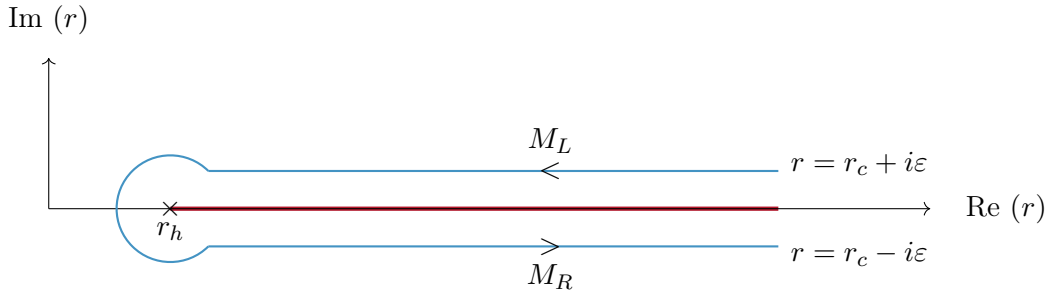
\begin{figure}[ht]
\begin{center}
\begin{tikzpicture}
\draw[red,ultra thick] (0,0) -- (8,0);
\node at (11,0) {$ \text{Re}\  (r) $};
\draw[black,->](-1.6,0) -- (10,0);
\draw[black,->](-1.6,0) -- (-1.6,2);
\node at (-1.6,2.5) {$ \text{Im}\  (r) $};
\node at (0,0) {$\times$};
\draw[blue, thick] (8,0.5) -- (0.5,0.5);
\draw[blue, thick] (0.5,0.5) arc (45:315:0.707);
\draw[blue, thick] (8,-0.5) -- (0.5,-0.5);
\node at (5,0.5) {$<$};
\node at (5,0.9) {$ {\scriptsize{M_L}}$};
\node at (5,-0.5) {$>$};
\node at (5,-0.9) {$M_R$};
\node at (0,-0.3) {$r_h$};
\node at (8.3,0.6) {$\qquad \qquad r=r_c+i\varepsilon $};
\node at (8.3,-0.6) {$\qquad \qquad r=r_c-i\varepsilon $};
\end{tikzpicture}
\end{center} 
\caption{Picture of bulk at a fixed $v$: red line is a branch cut from horizon to boundary}
\label{bulk contour}
\end{figure}

In \cite{Glorioso:2018mmw} the authors propose to fill in this boundary time contour by a complexified doubled bulk space-time as shown in Fig \ref{bulk contour}.
Consider the Schwarzschild AdS$_{d+1}$ black brane 
written in ingoing Eddington-Finkelstein coordinates that are regular at the future horizon,
\begin{equation}\label{eq:efads1}
ds^2 = -r^2\, \left(1 - \frac{r_h^d}{r^d}\right) \, dv^2 + 2\, dv\, dr + r^2\, d{\bf x}^2 \, . 
\end{equation}	
The coordinate $v$ becomes the time coordinate $t$ on the boundary of the spacetime at $r \to \infty$. One has to consider two copies of the black brane in the bulk that have to be stitched in a certain way across their future horizon \cite{Glorioso:2018mmw}. The proposed bulk spacetime correctly asymptotes to the aforementioned Schwinger-Keldysh time contour. 
In the bulk the radial coordinate has to be complexified to handle the near horizon region as proposed in \cite{Glorioso:2018mmw}. Define a complex radial tortoise coordinate $\zeta$
\begin{equation}\label{eq:ctordef2}
\frac{dr}{d\zeta} = \frac{i\,\beta}{2} \, r^2\, \left(1 - \frac{r_h^d}{r^d}\right) \,,
\end{equation}	
where $\beta = \frac{4\pi}{d \,r_h}$ is also the inverse Hawking temperature of the black hole.
$\zeta$ has a logarithmic branch point at $r=r_h$ coming from the integral of the blackening factor (about its zero). The branch cut in $\zeta$ is taken from $r=r_h$ to $r=\infty$. 
The jump of real part of $\zeta$ across the horizon is 1. 
Hence $\zeta$ parametrizes a double-sheeted spacetime that describes the bulk extension of the boundary Schwinger-Keldysh contour. 
On each sheet imaginary part  of $\zeta$ runs from $0$ at the AdS boundary to $\infty$ at the horizon. The real part of $\zeta$ changes between the two sheets and is given by the monodromy around the horizon. The real part of $\zeta$ is chosen to be zero on the left sheet $M_L$, then the real part of $\zeta$ becomes one on the right sheet $M_R$ after picking up the monodromy at the horizon. We will also have a UV cut-off of the geometry at $r=r_c$ for computational purposes. Hence asymptotically
\begin{equation}\label{eq:ctorbc1}
\zeta(r_c+i\,\varepsilon) = 0 \,, \qquad \zeta(r_c-i\,\varepsilon) = 1\,.
\end{equation}	
The metric written in $\zeta$ coordinate becomes
\begin{equation}\label{eq:sadsct1}
ds^2 = -r^2\, \left(1 - \frac{r_h^d}{r^d}\right) \, dv^2 +  i\, \beta\, r^2 \, \left(1 - \frac{r_h^d}{r^d}\right)\,  dv\, d\zeta + r^2\, d{\bf x}^2 \,, 
\end{equation}	
where $r(\zeta)$ is obtained by integrating \eqref{eq:ctordef2}. 
The complex tortoise coordinate for Schwarzschild-AdS$_{d+1}$ geometry becomes \cite{Jana:2020vyx}
\begin{equation}\label{eq:ctoradsd}
 \zeta +\zeta_c=  
\frac{i\, d}{2\pi\, (d-1)}  \left( \frac{r}{r_h}\right)^{d-1}\, {}_2F_1\left(1,\frac{d-1}{d};2-\frac{1}{d};\frac{r^d}{r_h^d}\right) ,
\end{equation}	
where $\zeta_c$ is considered to make $\zeta=0$ at $r=r_c+i\,\varepsilon$. The branch-cut of the hypergeometric function runs from $r=r_h$ to $r=\infty$. 
With this set up, one has to consider all bulk fields to live on a complex $\zeta$ space and think of the classical bulk action to be a contour integral over the complex tortoise coordinate $\zeta$. Hence
\begin{equation}\label{eq:}
S_\text{bulk} = \oint d\zeta \int d^dx \, \sqrt{-g} \; \mathcal{L}[g_{AB}, \Phi]\,
\end{equation}	
where $x^\mu$ are the boundary coordinates. This action serves as the starting point for computations of influence functionals. 
\par
\begin{figure}[ht]
\begin{center}
\begin{tikzpicture}
\begin{feynman}
\draw[red,ultra thick] (0,0) -- (4,0);
\draw[red,ultra thick] (-2,0) -- (-4,0.5);
\node at (-2,-0.3) {$r_-$};
\node at (0,0) {$\times$};
\draw[blue, thick] (4,0.5) -- (0.5,0.5);
\draw[blue, thick] (0.5,0.5) arc (45:315:0.707);
\draw[blue, thick] (4,-0.5) -- (0.5,-0.5);
\node at (2,0.5) {$<$};
\node at (2,0.9) {$ {\scriptsize{M_L}}$};
\node at (2,-0.5) {$>$};
\node at (2,-0.9) {$M_R$};
\node at (0,-0.3) {$r_+$};
\node at (4.3,0.6) {$\qquad \qquad \qquad \qquad \qquad r=r_c+i\varepsilon,\chi=0, \Theta=0 $};
\node at (4.3,-0.6) {$\qquad \qquad \qquad \qquad \qquad r=r_c-i\varepsilon, \chi=1, \Theta=\mu_+ $};
\end{feynman}
\end{tikzpicture}
\caption{Picture of bulk at a fixed $v$: red line is branch cut from horizon to boundary}
\label{fig2}
\end{center} 
\end{figure}
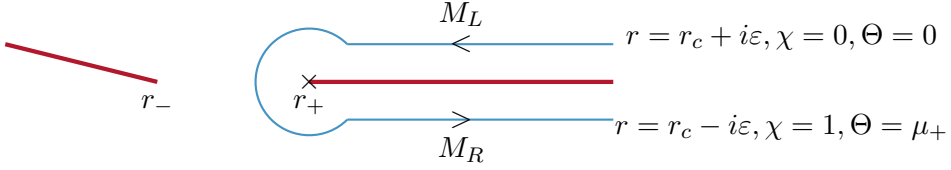

 
In the following we extend the holographic contour mentioned above for rotating BTZ black holes with outer and inner horizons. The two horizons correspond to two branch points of the complexified tortoise coordinate in this case. Let us start by writing the rotating BTZ black hole metric in standard coordinates \cite{Son:2002sd}
\begin{eqnarray}
 \mathrm{d}s^2 = -f \mathrm{d}t^2 + \frac{\mathrm{d}r^2}{f}+r^2 \bigl( \mathrm{d}\phi -  n_{\phi}\, \mathrm{d}t \bigr)^2\,; \ f(r) = \frac{(r^2-r_+^2)(r^2-r_-^2)}{\, r^2}\,,\  n_{\phi}(r)=\frac{r_+
r_-}{ \,r^2}\,
\label{BTZmetric}
\end{eqnarray}
where $ n_{\phi}(r)$ is the angular velocity of the black hole at radius r. We will denote the two angular velocities at the two horizons to be $ n_{\phi}(r_+)=\frac{r_{-}}{r_+}=\mu_+$  and $ n_{\phi}(r_{-})=\frac{r_+}{r_{-}}=\mu_-$. 
The two inverse temperatures associated to the two horizons are denoted by  $\beta_+$ and $\beta_-$ where
\begin{align}
\beta_\pm=\frac{2\pi r_\pm}{r_+^2-r_-^2}\ .
\end{align}
The surface gravities $\kappa_{\pm}$ of the two horizons are
\begin{align}
\kappa_{\pm}=\frac{r_+^2-r_-^2}{r_{\pm}}\ .
\end{align}
Let us now introduce the ingoing Eddington-Finkelstein coordinates for the rotating BTZ metric that are regular at the future horizon
\begin{subequations}
 \begin{align}
  \mathrm{d}v= \mathrm{d}t+\frac{\mathrm{d}r}{f}\ , \ \mathrm{d}\phi'= \mathrm{d}\phi +\frac{ n_{\phi}}{f}\mathrm{d}r\ .
  \end{align}
  \end{subequations}
The radial tortoise coordinate $r^*$ is defined by 
$dr^*=\frac{dr}{ f(r)}$.
A closed form expression of $r^*$ is 
\begin{align}
r^*=\frac{1}{2(r_+^2-r_-^2)}\left\{r_+\log\left(\frac{r-r_+}{r+r_+}\right)+r_-\log\left(\frac{r+r_-}{r-r_-}\right)\right\}\ .
\end{align}
The tortoise coordinate $r^*$ maps $[r_+,\infty)$ to $(-\infty,0)$. Let us also define another coordinate $r^\#$ by
 $ \mathrm{d}r^{\#}= \frac{ n_{\phi}}{f}\mathrm{d}r$.
  $r^\#$ also runs from $-\infty$ to $0$ as one moves from horizon to infinity.
 Explicitly the coordinate $r^{\#}$ is
\begin{align}
r^{\#}
    &=\frac{1}{2(r_+^2-r_-^2)} \bigg[r_- \ln \bigg(\frac{r-r_+}{r+r_+}\bigg)+r_+ \ln \bigg(\frac{r+r_-}{r-r_-}\bigg)\bigg]\ .
    \label{rhash}
\end{align}
In terms of these coordinates the ingoing Eddington-Finkelstein coordinates become 
\begin{subequations}
  \begin{align}
    v =t+r^*\ , \ \phi'=\phi+r^{\#}\ .
  \end{align}
  \end{subequations}
  Let us now define a complexified tortoise coordinate $\chi$ (a rescaling of $r^*$) by
\begin{align}
  \chi(r)&= \oint_{r_c+i\varepsilon }^{r}\frac{dr'}{\frac{i\beta }{2}f(r')}\nonumber\\
 &=\frac{1}{2\pi i}\log\left\{\left(\frac{r-r_+}{r+r_+}\right)\left(\frac{r+r_-}{r-r_-}\right)^{\frac{r_-}{r_+}}\left(\frac{r_c+r_+}{r_c-r_+}\right)\left(\frac{r_c-r_-}{r_c+r_-}\right)^{\frac{r_-}{r_+}}\right\}\ .
\label{eq:chi}
\end{align}
Note that the normalisation is chosen in such a way that $\chi(r_c+i\varepsilon)=0$. $ \chi(r)$ has branch points at $r_+$ and $r_-$. We will consider branch cut in $\chi$ that extend from $r=r_+$ to $r=\infty$. 
The inner horizon branch cut is chosen such that $\chi$ is analytic in the real interval between the two horizons. 
Since we will be interested in the region outside the future horizon $r_+$, we will now consider the complexified doubled bulk spacetime as shown in Fig \ref{fig2}. 

We will glue two copies of rotating BTZ black holes across their future horizon by a horizon cap and the monodromy in $\chi$ across the branch point at $r=r_+$ will be one. Hence
\begin{equation}\label{eq:ctorbc1}
\chi(r_c+i\,\varepsilon) = 0 \,, \qquad \chi(r_c-i\,\varepsilon) = 1\,.
\end{equation}
Let us also consider the following contour integral 
\begin{align}
\Theta(r)&=\oint  n_{\phi} d\chi=\frac{2}{i\beta}\oint dr\frac{ n_{\phi}}{f}\nonumber\\
&
=\frac{1}{2\pi i}\log\left\{\left(\frac{r+r_-}{r-r_-}\right)\left(\frac{r-r_+}{r+r_+}\right)^{\frac{r_-}{r_+}}\left(\frac{r_c-r_-}{r_c+r_-}\right)\left(\frac{r_c+r_+}{r_c-r_+}\right)^{\frac{r_-}{r_+}}\right\}\ .
\label{eq:theta}
\end{align}
Note that $\Theta(r)$ goes from $0$ to $\mu_+$ in going from $r_c+i\epsilon$ to $r_c-i\epsilon$ by picking up a monodromy across the branch point at $r=r_+$, giving
\begin{equation}\label{eq:ctorbc1}
\Theta(r_c+i\,\varepsilon) = 0 \,, \qquad \Theta(r_c-i\,\varepsilon) = \mu_+\,.
\end{equation}
This explains the origin of the chemical potential $\mu_+$ arising from the holographic contour in a rotating BTZ black hole.
The metric rewritten in ingoing $(v,\chi,\phi')$ coordinates is given by
\begin{align}
 ds^2&=- f dv^2+i \beta  f dv d\chi+r^2(d\phi^{\prime}-n_{\phi} dv)^2\ .
  \label{vchiphi'}
 \end{align}
\par
\begin{figure}[ht]
\begin{center}
\begin{tikzpicture}
\begin{scope}[thick]
\draw [black,->](0,0) --(0,2);
\draw [black,->](0,0) --(3.5,0);
\draw [black,->](0,0) --(-1.8,-1.8);
\draw [black](0,-0.2) --(-0.4,-1.5);
\draw [black,dashed](-0.4,-0.4) --(-0.4,-1.5);
\draw [black,dashed](0,-0.2) --(-0.,-1.1);
\draw [black,dashed](-0.4,-1.5) --(-0.,-1.1);
\filldraw[color=red!60, fill=red!60, very thick](-0.4,-1.5) circle (.1);
\filldraw[color=red!60, fill=red!60, very thick](-0.0,-0.0) circle (.1);
\draw [magenta](0,0.1) --(3.0,0.1);
\draw [magenta](3.0,0.1) --(3.0,-0.2);
\draw [magenta](0.0,-0.2) --(3.0,-0.2);
\draw [black](1.5,0) --(0,0);
\end{scope}
\node at (0.,2.2) {$\text{Im}(t)$}; 
\node at (4.0,0.0) {$\text{Re}(t)$};
\node at (-1.2,-0.6) {$\text{Im}(\phi)$};
\node at (3.0,-0.4) {};
\end{tikzpicture}
\end{center}
\caption{SK contour with an initial state at finite temperature and chemical potential.}
\label{chemcontour}
\end{figure}
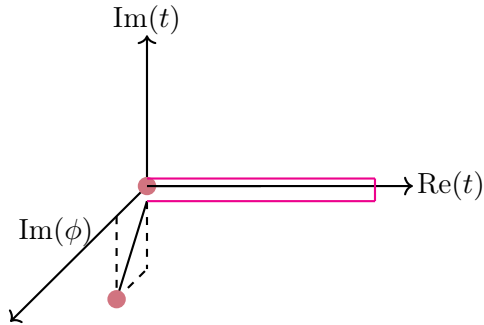

The CFT contour for a given initial state at finite temperature and chemical potential for angular momentum in this case is described as in Fig \ref{chemcontour} (following \cite{Skenderis:2008dg}). In \cite{Skenderis:2008dg} the authors described how the contour for the boundary CFT corresponding to a rotating black hole does not only lie in the complex t plane, but also has a leg into the complex $\phi$ plane. The solid circles in Fig \ref{chemcontour} are to be identified. In this paper we are interested in computing the generating functional for CFT correlators holographically for a given initial state at finite temperature and chemical potential due to angular momentum. 

\section {Probe scalar dynamics in bulk}
\label{sec:Probe scalar}
\subsection{Ingoing solution}
In this section we consider a free massive probe scalar field $\Phi$ with mass $M$ minimally coupled to the rotating BTZ black hole. According to AdS/CFT correspondence, the scalar field in the bulk is dual to a scalar operator $\mathcal{O}$ in the boundary CFT with conformal dimension $\Delta=1+\sqrt{1+M^2}$. 

The action for the probe scalar field  is given by
\begin{align}
  S&= -\oint d\chi dv d\varphi '\sqrt{-g}\left[ \frac{1}{2}g^{A B}\partial_{A}\Phi \partial_{B}\Phi  +\frac{M^2}{2}\Phi^2\right]\ .
  \label{eq1}
  \end{align}
  The classical scalar equation of motion is the Klein-Gordon equation
  \begin{align}
\frac{1}{\sqrt{-g}}\partial_A[\sqrt{-g}g^{AB}\partial_B \Phi]-M^2\Phi=0\ .
\end{align}
In the following we solve the massive scalar equation with appropriate boundary conditions to find the ingoing bulk to boundary green's function. 
To solve the massive scalar equation, it is useful to define a new radial coordinate $z$ as in \cite{Dias:2019ery}
\begin{align}
z= \frac{r^2-r_-^2}{r_+^2-r_-^2}\ .
 \label{zcoord}
\end{align}
In terms of the $z$ coordinate, the inner (Cauchy) horizon at $r=r_-$ is located at $z=0$, the event horizon at $r=r_+$ is at $z=1$ and the asymptotic boundary is at $z=+\infty$.
We will consider a solution of the scalar equation of the form
\begin{equation}
\Phi_{in} = \int \frac{ d \omega}{(2\pi)}\sum_{m} G^+( \omega,z,m)e^{i(m\phi^{\prime}- \omega v)}\,.
\label{ansatz}
\end{equation}
where $G^{+}( \omega,z,m)$  is the frequency domain retarded (ingoing) bulk to boundary Green's function. $\omega$ and $m$ are frequency and azimuthal number of the mode. 
 Written in z coordinate the equation of motion looks as following
\begin{align}
&z (z-1)\partial^2_z G^+ + \bigg[2z-1-\frac{i \omega \{z(r_+^2-r_-^2)+r_-^2\}^{1/2}}{r_+^2-r_-^2}+\frac{imr_+ r_- }{(r_+^2-r_-^2)\{z(r_+^2-r_-^2)+r_-^2\}^{1/2}}\bigg]\partial_z G^+\nonumber\\
& -\frac{1}{4 \{z(r_+^2-r_-^2)+r_-^2\}^{1/2}}\bigg[i \omega+ \frac{imr_+ r_- }{\{z(r_+^2-r_-^2)+r_-^2\}}+\frac{m^2}{\{z(r_+^2-r_-^2)+r_-^2\}^{1/2}}\bigg]G^+ -\frac{M^2}{4}G^+=0 \,.
\end{align}
If we consider an ansatz of the form 
\begin{align}
  G(z)&=e^{i(\omega r^*- m r^{\#})}z^{\alpha}(1-z)^{\beta}F(z)
   \end{align} then the equation of motion in terms of F(z) becomes a hypergeometric differential equation. 
We impose regularity at the horizon and normalisability at the cut-off boundary to solve the equation of motion. The boundary conditions are given by
\begin{align}
  \frac{\mathrm{d}G^+(\omega,r,m)}{\mathrm{d}\chi}\bigg\rvert_{r=r_+}=0\ , \quad G^+\left(  \omega ,r _c ,m \right)= 1\ .
  \label{boundarycond}
\end{align}
After using hypergeometric identities, the regular solution that satisfies the above boundary conditions becomes (after writing the mass $M$ of the scalar in terms of $\Delta$) (similar solution appears in a very recent paper \cite{Natsuume:2020snz})
\begin{align}
G^+(z)=  \frac{e^{i\left( \omega r^*-m r^{\#} \right)}z^{\frac{c-1}{2}-a+\frac{\Delta}{2}}(z-1)^{\frac{a+b-c}{2}}{}_2F_1\left( a-c+1-\frac{\Delta}{2},a-\frac{\Delta}{2},a+b-c+1; 1-\frac{1}{z} \right)}{e^{i\left( \omega r_c^*-m r_c^{\#} \right)}z_c^{\frac{c-1}{2}-a+\frac{\Delta}{2}}(z_c-1)^{\frac{a+b-c}{2}}{}_2F_1\left( a-c+1-\frac{\Delta}{2},a-\frac{\Delta}{2},a+b-c+1; 1-\frac{1}{z_c} \right)}
\label{solnn}
\end{align}
where
 \begin{subequations}
 \begin{align}
    a&= \frac{1}{2}\bigg(2-\frac{i( \omega-m \mu_-)}{\kappa_-}-\frac{i( \omega-m \mu_+)}{\kappa_+}\bigg)\\
    b&= \frac{1}{2}\bigg(\frac{-i( \omega-m \mu_-)}{\kappa_-}-\frac{i( \omega-m \mu_+)}{\kappa_+}\bigg)\\
    c&= 1-\frac{i( \omega-m \mu_-)}{\kappa_-}\ .
   \end{align}
      \label{def}
   \end{subequations}   
   This is the exact ingoing bulk to boundary green's function that corresponds to the infalling quasi normal modes of the black hole. The ingoing solution is analytic in $z$ coordinate. 
\subsection{Solution on the SK contour}
In this section, first we study the dynamics of the outgoing Hawking modes. Within the holographic Schwinger-Keldysh path integral framework the outgoing modes naturally arise as frequency and angular momentum reversed counter parts of the ingoing modes. Hence
the physics of the outgoing Hawking radiation is captured by the advanced bulk to boundary green's function. In the following we will construct the advanced green's function by imposing (generalised) time-reversal transformation on the ingoing solution. 
Under (generalised) time reversal transformation the coordinates and parameters transform as follows
\begin{subequations}
\begin{align}
v&\rightarrow i\beta\chi-v\\
\chi &\rightarrow \chi\\
\phi' &\rightarrow i\beta\Theta -\phi'\\
 \omega&\rightarrow - \omega\\
m&\rightarrow -m\ .
\end{align}
\label{coordreversal}
\end{subequations}
These set of transformations keep the metric in equation \eqref{vchiphi'} manifestly time reversal invariant once the metric is written in the following way 
\begin{align}
ds^2=-fdv\left(dv-i\beta d\chi\right)-r^2\left(d\phi'- n_{\phi} dv\right)\left( i n_{\phi} \beta d\chi-d\phi'+ n_{\phi}\left(dv-i\beta d\chi\right)\right)\ .
\end{align}
Under time reversal, the ingoing solution $\Phi_{in}$ in equation \eqref{ansatz} gets mapped to the outgoing solution $\Phi_{out}$
\begin{subequations}
\begin{align}
\Phi_{out}(v,z,\phi')&=\int\frac{d \omega}{2\pi}\sum_m e^{-i \omega v+i m\phi'}G(- \omega,z,-m)e^{-\beta\left( \omega\chi-m\Theta\right)}\ .
\end{align}
\end{subequations}
Note that the outgoing solution is non-analytic in $z$ coordinate because of the presence of $\chi$ and $\Theta$ in the exponential factor. $\chi$ and $\Theta$ have logarithmic branch point at $r=r_+$.
\par
Now we construct the full solution on the doubled bulk spacetime by adding the ingoing and outgoing solutions. The full solution captures the ingoing quasi normal modes as well as the outgoing Hawking modes. 
The full solution on the holographic SK contour is given by the following linear combination of the ingoing and outgoing solutions
\begin{align}
  \Phi &= \int \frac{d \omega }{2\pi }\frac{1}{2\pi }\sum_{m}\left\{ c_{ \omega ,m}\ G^+\left(  \omega ,z,m \right)+h_{\omega ,m}\  G^+\left( - \omega ,z,-m \right)e^{-\beta \left(  \omega \chi -m\Theta  \right)} \right\}e^{-i  \omega v+im \phi'}\ .
  \label{ansatz2}
\end{align}
At the boundary the doubled bulk spacetime asymptotes to CFT Schwinger-Keldysh contour with left and right sources $\phi_L$ and $\phi_R$. 
 Hence from the requirement $\Phi \left( r_c+ i\varepsilon  \right)=\phi_L$ and $\Phi \left(r_c- i\varepsilon \right)=\phi_R$, we find that
\begin{subequations}
\begin{align}
c_{ \omega ,m}
  &= -f_{\omega ,m}\ \phi_L\left(  \omega ,m \right)+\left( 1+f_{\omega ,m} \right)\phi_R\left(  \omega ,m \right)\nonumber\\
h_{\omega ,m}  &= \left( 1+f_{\omega ,m} \right)\left( \phi_L\left( \omega ,m \right)-\phi_R\left(  \omega ,m \right) \right)
    \label{coefficients}
\end{align}
\end{subequations}
where $f_{\omega, m}= \frac{1}{e^{\beta(\omega-m \mu_+)}-1}$ is the Bose-Einstein factor with chemical potential.
Hence the full solution of the scalar wave equation in equation \eqref{ansatz2} becomes
\begin{align}
\begin{split}
  \Phi &= \int \frac{d \omega }{2\pi }\frac{1}{2\pi }\sum_m\left\{ \left( -f_{\omega ,m}\ \phi_L\left(  \omega ,m \right)+\left( 1+f_{\omega ,m} \right)\phi_R\left(  \omega ,m \right) \right)G^+\left(  \omega ,z,m \right)\right.\\
  &\left.+\left( 1+f_{\omega ,m}\right)\left( \phi_L\left(  \omega ,m \right)-\phi_R\left(  \omega ,m \right) \right)G^+\left( - \omega ,z,-m \right) e^{-\beta \left(  \omega \chi -m\Theta  \right)} \right\}e^{-i \omega v+im \phi'}\ .
  \end{split}
  \label{ansatz3}
\end{align}
We will use this exact solution to compute the quadratic influence functional in section \ref{quad1}. 
\section {Influence functional of open scalar field}
\label{sec:Influence functional}
In this section we compute the quadratic and higher order influence functionals of the open scalar field. 
To compute the influence functional it is useful to work with the retarded-advance (RA) basis \cite{ chaudhuri2019spectral, aurenche1992comparison, van1994transformations} within the SK framework where the KMS relations become manifest. The RA basis is defined as
\begin{subequations}
\begin{align}
  \phi_F\left(  \omega ,m \right)&= -\left( 1+f_{\omega ,m} \right)\phi_R\left(  \omega ,m \right)+f_{\omega ,m}\phi_L\left(  \omega ,m \right)\\
  \phi_P\left(  \omega ,m \right)&= -f_{\omega ,m}\left( \phi_R\left(  \omega ,m \right)-\phi_L\left(  \omega ,m \right) \right)\ .
  \label{R.Abasis}
\end{align}
\end{subequations}
 In terms of the RA basis we can write down the scalar solution given in equation \eqref{ansatz3} as
\begin{align}
  \Phi&= \int \frac{d \omega }{2\pi }\frac{1}{2\pi }\sum_m\left\{ -G^+\left( \omega ,z,m  \right)\phi_F+G^+\left( - \omega ,z,-m \right)\phi_P e^{\beta \left(  \omega \left( 1-\chi  \right)-m\left( \mu_+ -\Theta  \right) \right)} \right\} e^{-i \omega v+im \phi '}\ .
  \label{ansatz4}
\end{align}
Using this scalar solution in RA basis, we compute the quadratic on-shell action in section \ref{quad1}.
\subsection{Quadratic effective theory}
\label{quad1}
Imposing the Klein-Gordon equation of motion, the quadratic on-shell action of the scalar field becomes
\begin{align}
  S^{on-shell}
  &= -\oint d\chi dv d\varphi '\left\{ \partial_{A}\left( \frac{\sqrt{-g}}{2}g^{A B}\Phi \partial_{B}\Phi  \right)\right\}\nonumber\\
  &= -\oint d\chi dv d\phi '\left\{ \partial_{\chi }\left( \frac{r}{2}\Phi \partial_{v}\Phi  \right)+\partial_{\chi }\left( \frac{-ir }{\beta }\Phi \partial_{\chi }\Phi  \right)+\partial_{\chi }\left( \frac{ n_{\phi} r}{2}\Phi \partial_{\phi '}\Phi  \right)\right\}\nonumber\\
  &= -\int dv d\phi ' \left[ \frac{r}{2}\Phi \partial_{v}\Phi -\frac{ir}{\beta }\Phi \partial_{\chi }\Phi +\frac{ n_{\phi} r}{2}\Phi \partial_{\phi' }\Phi  \right]\bigg\rvert^{\chi =1,\Theta=\mu_+}_{\chi =0,\Theta=0}\ .
  \label{on-shell2}
\end{align}
In the above equation we see that the quadratic on-shell action is a total derivative, hence it has to be evaluated at the boundary.  
In RA basis the quadratic influence functional becomes (see appendix \ref{sec:integrals1} for details)
\begin{align}
S^{on-shell}= \int\frac{d\omega }{2\pi }\frac{1}{2\pi }\sum_m  \phi_F(\omega,m)\phi_P(-\omega,-m)\mathcal{G}_{FP}[\omega,m]
\label{q11}
\end{align}
where after setting $G^{+}(\omega,z_c,m)=1$, 
\begin{align}
\mathcal{G}_{FP}[\omega, m]&=-\frac{2ir_c}{\beta}\left[\partial_{\chi}G^+(\omega,z_c,m)+\frac{\beta}{2}(\omega-m n_{\phi})\right](1-e^{-\beta(\omega-\mu_+ m)})\ .
\end{align}
The FF and PP terms do not contribute to the radial integral since they are purely analytic functions (the $\chi$ dependence goes away inside the integral) and contributions from the left and right contours cancel each other. We will compute the quadratic influence functional by substituting $G^+$ in the above expression. Before proceeding with the calculation let us define yet another basis known as the average-difference/Keldysh basis in which the fluctuation-dissipation relation will be studied. This basis will be more useful in the computation of higher order influence phase that will be computed in a small frequency and momentum expansion. We define the average-difference basis \cite{kamenev_2011, Keldysh:1964ud, CHOU19851} in the following 
\begin{subequations}
\begin{align}
  \phi_F\left(  \omega ,m \right)&= -\phi_a\left(  \omega ,m \right)-N_{\omega,m}\phi_d\left(  \omega ,m \right)\\
  \phi_P\left(  \omega ,m \right)&= -\phi_d\left(  \omega ,m \right)f_{ \omega ,m }
  \label{a-dbasis}
\end{align}
\end{subequations}
where
$N_{\omega,m}= f_{\omega,m}+\frac{1}{2}$\ . In terms of  the left and right sources $\phi_L$ and $\phi_R$ at the boundary, the average source (mean value) becomes $\phi_a=\frac{\phi_R+\phi_L}{2}$ and the difference source (quantum/statistical fluctuations) becomes $\phi_d=\phi_R-\phi_L$. The quadratic influence functional of equation \eqref{q11} written in average-difference basis becomes
\begin{align}
  S^{on-shell}&= \int \frac{d \omega }{2\pi } \sum_m\left( \mathcal{G}_{ad}\phi_a\left(  \omega ,m \right)\phi_d\left(- \omega ,-m  \right)+\mathcal{G}_{dd}\phi_d\left(  \omega ,m \right)\phi_d\left( - \omega ,-m \right) \right)
  \label{on-shellad}
\end{align}
where the coefficient functions that appear in front of the $\phi_a \phi_d$ and $\phi_d \phi_d$ terms are 
\begin{align}
\begin{split}
  \mathcal{G}_{ad}
  &= 2 \frac{ir_c}{\beta } \left[\partial_{\chi}G^+(\omega,z_c,m)+\frac{\beta}{2}(\omega-m n_{\phi})\right]\ ,
  \end{split}\\
  \begin{split}
  \mathcal{G}_{dd} &= \frac{ir_c}{2\beta }\frac{e^{\beta \left(  \omega -m\mu_+  \right)}+1}{e^{\beta \left(  \omega -m\mu_+  \right)}-1}\left[\partial_{\chi}G^+(\omega,z_c,m)+\frac{\beta}{2}(\omega-m n_{\phi})\right]+(\omega\rightarrow -\omega)\ .
  \end{split}
  \label{coefficients-ad}
\end{align}
where we have written $\mathcal{G}_{dd}$ in a symmetric way under $\omega\rightarrow -\omega$.
Note that there is no $\mathcal{G}_{aa}$ term in the quadratic influence functional. This is a consequence of the microscopic unitarity of the theory. 
After substituting for the solution of $G^+\left( \omega ,z_c,m \right)$ from eq. \eqref{solnn} and picking up the coefficient of $r_c^{4-2\Delta}$ we obtain the coefficient functions $\mathcal{G}_{ad}$ and $\mathcal{G}_{dd}$. Note that this will be the result after an appropriate counter term subtraction as is done in holographic renormalisation. We finally obtain 
\begin{subequations}
\begin{align}
  \mathcal{G}_{ad}[\omega ,m]&=  -\frac{2\left(r_+^2-r_-^2\right)^{\Delta-1} \Gamma^2 \left( 2-\Delta  \right)\Gamma (\mathfrak{p}_+)\Gamma (\mathfrak{p}_-)\sin\left( \pi \Delta  \right)}{\pi \Gamma \left( 1-\Delta +\mathfrak{p}_+ \right)\Gamma \left( 1-\Delta +\mathfrak{p}_- \right)}\\
  &= -\frac{2\left( r_+^2-r_-^2 \right)^{\Delta-1} |\Gamma (\mathfrak{p}_+)\Gamma (\mathfrak{p}_-)|^2\sin\left( \pi (\Delta -\mathfrak{p}_+) \right)\sin\left( \pi (\Delta -\mathfrak{p}_-) \right)}{\pi \Gamma^2 (\Delta -1)\sin(\pi \Delta )}\\
  &= -\frac{2\left( r_+^2-r_-^2 \right)^{\Delta-1} |\Gamma (\mathfrak{p}_+)\Gamma (\mathfrak{p}_-)|^2\left( \cosh\left( \frac{\beta _-\omega _-}{2} \right)-\cosh\left( \frac{\beta _+\omega _+}{2}-i\pi \Delta  \right) \right)}{2\pi \Gamma^2 (\Delta -1)\sin\left( \pi \Delta  \right)}\\
  &=-\frac{\left( r_+^2-r_-^2 \right)^{\Delta-1} |\Gamma (\mathfrak{p}_+)\Gamma (\mathfrak{p}_-)|^2}{\pi \Gamma^2(\Delta -1)\sin\left( \pi \Delta  \right)}\\
 & \left(\cosh\left(\frac{\beta_-\omega_-}{2}\right)
  -\cosh\left(\frac{\beta_+\omega_+}{2}\right)\cos(\pi\Delta)+i\sinh\left(\frac{\beta_+\omega_+}{2}\right)\sin(\pi\Delta)\right)
  \label{rinfluencephase}
\end{align}
\end{subequations} where various parameters that appear in the above expression are defined as follows
\begin{subequations}
\begin{align}
  \mathfrak{p}_+ &= \frac{\Delta }{2}+i\frac{(m-\omega )}{2(r_+-r_-)}\\
  \mathfrak{p}_- &= \frac{\Delta }{2}-i\frac{(m+\omega )}{2(r_+ +r_-)}\\
  \omega_+&=\omega-m \mu_+=\omega-m \frac{r_-}{r_+}\\
  \omega_-&=\omega-m\mu_-=\omega-m\frac{r_+}{r_-}\\
  \beta_+ &=\frac{2\pi r_+}{r_+^2-r_-^2}\\
  \beta_- &=\frac{2\pi r_-}{r_+^2-r_-^2}\ .
  \label{pppmdef}
\end{align}
\end{subequations}
$ \mathfrak{p}_+$, $ \mathfrak{p}_-$, $\omega_+$ and $\omega_-$ are lightcone like dimensionless combination of frequency and momentum. 
From a similar computation of $\mathcal{G}_{dd}$ we get
\begin{align}
\mathcal{G}_{dd}[\omega,m]&=\frac{-i}{2}\frac{\left( r_+^2-r_-^2 \right)^{\Delta-1} |\Gamma (\mathfrak{p}_+)\Gamma (\mathfrak{p}_-)|^2}{\pi \Gamma^2(\Delta -1)}\cosh\left(\frac{\beta_+\omega_+}{2}\right)\ .
\label{bc1}
\end{align}
By comparing equation \eqref{bc1} with equation \eqref{rinfluencephase} 
we get the fluctuation-dissipation relation in presence of chemical potential due to angular momentum given by
\begin{align}
\mathcal{G}_{dd}=\frac{i}{2}\coth\left(\frac{\beta\left( \omega-m\mu_+\right)}{2}\right)Im(\mathcal{G}_{ad})\ .
\label{fluct-dissip3}
\end{align}
$Im(\mathcal{G}_{ad})$ gives the quadratic spectral function at finite temperature \cite{CHOU19851, Haehl:2017eob}. 
\par Our quadratic influence functional results computed using a natural extension of the recently proposed holographic SK prescription of \cite{Glorioso:2018mmw} match with the long known results of \cite{Son:2002sd} under following coordinate transformations and redefinition of parameters
\begin{subequations}
\begin{align}
  x^+ &= r_+ t- r_- \phi \ , \ x^- = -r_- t+ r_+ \phi \ , \\
  k_+ &= \frac{\omega r_+ -m r_-}{r_+^2-r_-^2}=\frac{\beta _+\omega _+}{2\pi } \\
  k_-&= \frac{\omega r_- -mr_+}{r_+^2-r_-^2}=\frac{\beta _-\omega _-}{2\pi } \\
  p_+ &= \pi T_L\left( k_+ +k_- \right)=\pi T_L\frac{(\omega -m)}{(r_+-r_-)}\\
  p_- &= \pi T_R\left( k_+ -k_- \right)=\pi T_R\frac{(\omega +m)}{(r_+ +r_-)}
\end{align}
\end{subequations}
where $x^\pm$ are comoving coordinates, $ k_\pm$ are momenta conjugate to $x_\pm$ and $T_L$ and $T_R$ are the left and right temperatures given by
\begin{align}
T_L=\frac{r_+-r_-}{2 \pi}\ , \ T_R=\frac{r_++r_-}{2 \pi}\ .
\end{align} 
The answer given in \cite{Son:2002sd} for $\mathcal{G}_{ad}$ goes as
\begin{align}
\begin{split}
  \mathcal{G}_{ad}[\omega ,m]&\sim \frac{|\Gamma \left(\frac{\Delta}{2} +i\frac{p_+}{2\pi T_L} \right)\Gamma \left( \frac{\Delta}{2} +i \frac{p_-}{2\pi T_R} \right)|^2}{\Gamma \left( 2\frac{\Delta}{2}-1 \right)^2\sin(2\pi \beta _+)}\left[ \cosh\left( \frac{p_+}{2T_L}-\frac{p_-}{2T_R} \right)\right.\\
  &\left.-\cos(2\pi \frac{\Delta}{2})\cosh\left( \frac{p_+}{2T_L}+\frac{p_-}{2T_R} \right)
  +i\sin\left( 2\pi \frac{\Delta}{2} \right)\sinh\left( \frac{p_+}{2T_L}+\frac{p_-}{2T_R} \right) \right]\ .
  \end{split}
  \label{sonstaricheck}
\end{align}

\subsection{Quartic  order effective theory}
\label{sub4}
In this section, we will turn on weak quartic self interaction of the massless scalar field. At the leading order in the self interaction strength this term corresponds to the four-point contact Witten diagram in the bulk. Going to frequency domain we determine the quartic order influence functional of the scalar field in a low frequency and angular momentum gradient expansion till first sub leading order in derivatives. We are relying on derivative expanded scalar solution to evaluate higher order influence phase rather than trying to compute them using exact scalar solution since the later might be technically somewhat complicated. 
\subsubsection{Quartic Influence functional}
In the following we add a quartic self-interaction term to the original action. We will compute the on-shell action by imposing the scalar equation of motion and integrating out the complexified radial coordinate. 
We consider a simpler setting where we have a massless self interacting scalar field in the bulk \footnote{The massless case is simpler since the scalar wave equation is simpler and solving it order by order in perturbation parameter is easier.} that is dual to a marginal CFT operator with conformal dimension two at the boundary. 
The action for the massless interacting scalar field is given by
\begin{align}
  S&= -\int d\chi dv\int d\phi' \sqrt{-g}\left[ \frac{1}{2}g^{A B}\partial_{A}\Phi\partial_{B}\Phi +\frac{\lambda }{4!}\Phi ^4 \right]
  \label{sfull}
\end{align}
where $\lambda$ is the interaction strength. 
The equation of motion of the scalar field resulting from the above action is
\begin{align}
  \frac{1}{\sqrt{-g}}\partial_{A}\left[ \sqrt{-g}g^{A B}\partial_{B}\Phi  \right]-\frac{\lambda }{3!}\Phi ^3&= 0\ .
  \label{EOM}
\end{align}
Since this equation is non-linear in $\Phi$, we will solve this equation perturbatively in $\lambda$.
We write the solution $\Phi$ as a series in $\lambda $
\begin{align}
  \Phi &= \sum_{n=0}^{\infty}\lambda ^n\Phi _n\ .
\end{align}
The action upto linear order in coupling constant $\lambda$ becomes
\begin{subequations}
\begin{align}
  S&=- \int dv\int d\phi' \oint d\chi \sqrt{-g}\left[ \frac{g^{A B}}{2}\left( \partial_{A}\Phi _0\partial_{B}\Phi _0+2\lambda \partial_{A}\Phi _1\partial_{B}\Phi _0 \right)+\frac{\lambda }{4!}\Phi _0^4 \right]\\
  \begin{split}
  &=- \int dv\int d\phi'\oint d\chi \sqrt{-g}\left[ \frac{1}{2}\partial_{A}\left( \sqrt{-g}g^{A B}\left( \Phi _0+2\lambda \Phi _1 \right)\partial_{B}\Phi _0 \right)\right.\\
  &\qquad \qquad \qquad \qquad \qquad \qquad \left.-\frac{1}{2}\left( \Phi _0+2\lambda \Phi _1 \right)\partial_{A}\left( \sqrt{-g}g^{A B}\partial_{B}\Phi _0 \right)+\frac{\lambda }{4!}\Phi _0^4 \right]
  \end{split}\\
 S_{on-shell} &=- \int dv \int d\phi'\oint d\chi \sqrt{-g}\left[ \frac{1}{2}\partial_{A}\left( \sqrt{-g}g^{A B}\Phi _0\partial_{B}\Phi _0 \right)+\frac{\lambda }{4!}\Phi _0^4 \right]\ .
\end{align}
\end{subequations}
In the last line we have imposed the e.o.m of the free scalar $\Phi _0$ and we choose $\Phi _1$ at the boundary to be zero as a choice of boundary condition. The first term in the above equation is the quadratic on-shell action that we have computed in the previous section. The quartic piece in the on-shell action is
\begin{align}
  S_{quartic, on-shell}&=- \int dv\int d\phi' \oint d\chi \sqrt{-g}\left[ \frac{\lambda }{4!}\Phi _0^4 \right]\ .
  \label{squartic0}
\end{align}
We have to perform the radial integral in order to obtain the quartic influence functional of the open scalar field at the boundary. To compute the quartic integral, we will first solve the free scalar field $\Phi _0$ in a gradient expansion in low frequency and angular momentum. We will then substitute this solution in the action and integrate out the radial coordinate. In the following we compute the derivative expansion of the free scalar field. 
\\
{\bf{Derivative expansion of scalar solution:}}\\
We will start by expanding the frequency domain bulk to boundary green's function in terms of $\beta\omega$ and $\beta m\mu$, both of them being dimensionless parameters. We have denoted $\mu_+$ as $\mu$, this is the notation we will follow now on. The derivative expansion is
\begin{align}
  G^+&= G_0^++\frac{\beta \omega }{2}G^+_{\omega }+\frac{\beta m\mu }{2}G^+_m+\left( \frac{\beta \omega }{2} \right)^2G^+_{\omega ^2}+\left( \frac{\beta m\mu }{2} \right)^2G^+_{m^2}+\left( \frac{\beta \omega }{2} \right)\left( \frac{\beta m\mu }{2} \right)G^+_{\omega m}+ . . .
  \label{Gderivativeexpansion}
\end{align} where $G^+_{\omega }$, $G^+_m$ and all the higher order pieces are only functions of the radial coordinate. 
In the following we determine the values of $G_0^+$, $G_{\omega }^+$ and $G_{m}^+$ for the free massless scalar field. Refer to appendix \ref{sec:derivative expansion} for derivative expansion of massive scalar field. 
We impose the following boundary conditions while writing the derivative expansion of the green's function. These boundary conditions correspond to regularity of the green's function at the horizon and its normalisability  at the UV cut-off $r_c$. The boundary conditions are
\begin{subequations}
\begin{align}
 \frac{dG_0^+}{d\chi }\rvert_{r_+}&= 0 \qquad G^+_0\rvert_{r_c}=1\\
\frac{d}{d\chi }G^+_{\omega ^n,m^l}&=0 \qquad G^+_{\omega ^n,m^l}\rvert_{r_c}= 0\hspace{1cm} \forall n,l>0.
   \label{boundarycond}
\end{align}
\end{subequations}
Eventually we will take $r_c\rightarrow \infty$ at the end of our computations. 
At the leading order in derivative expansion i.e. at $\mathcal{O}(\beta \omega)^0$ and $\mathcal{O}(\beta m \mu)^0$ the massless scalar e.o.m gives 
\begin{align}
  \partial_{\chi }\left( r\partial_{\chi }G^+_0 \right)&= 0\ .
\end{align}
 After imposing regularity at the horizon and normalisability at the cut-off boundary we get
\begin{align}
G^+_0=1\ .
\end{align}
Collecting $\mathcal{O}(\frac{\beta \omega }{2})$ terms in the scalar e.o.m we get 
\begin{align}
  \partial_{\chi }\left( r\partial_{\chi }G^+_{\omega } \right)+ \partial_{\chi }r &= 0\ .
\end{align}
We integrate the above equation and impose the two boundary conditions to get
\begin{align}
\begin{split}
G^+_{\omega}&=\frac{1}{i\pi}(1-\mu^2)\left\{-\frac{1}{2(\mu+1)}\log\left(\frac{r-r_-}{r_c-r_-}\right)+\frac{1}{2(\mu-1)}\log\left(\frac{r+r_-}{r_c+r_-}\right)\right.\\
&\left.\qquad \qquad\qquad \ -\frac{1}{\mu^2-1}\log\left(\frac{r+r_+}{r_c+r_+}\right)\right\}\ .
\label{eqq1}
\end{split}
\end{align}
Since we are derivative expanding the ingoing solution that is analytic in the radial variable, $G^+_{\omega}$ is also analytic between $r_+$ and $r_c$.
To find $G_m^+$, the required equation at $\mathcal{O}(\frac{\beta m \mu }{2})$ that we need to solve is
\begin{align}
  &\mu\partial_{\chi }\left( r\partial_{\chi }G^+_m \right)-\partial_{\chi }\left( \frac{r_+r_-}{r} \right)= 0\ .
\end{align}
Solving the above equation and imposing the boundary conditions we obtain
\begin{align}
 G_m^+&=\frac{1}{i\pi \mu}\left\{-\frac{(1-\mu)}{2}\log\left(\frac{r-r_-}{r_c-r_-}\right)+\frac{(\mu+1)}{2}\log\left(\frac{r+r_-}{r_c+r_-}\right)-\mu\log\left(\frac{r+r_+}{r_c+r_+}\right)\right\}\ .
 \label{eqq2}
\end{align}
$G_m^+$ is analytic between $r_+$ and $r_c$ as well for the same reason. Given the derivative expansion till linear order for the ingoing green's function, we obtain the full green's function by adding the ingoing and outgoing green's functions where the outgoing part is just a frequency and momentum reversed counter part of the derivative expanded ingoing solution. For computational ease we will collect the even and odd parts of the full Green's function under frequency and momentum reversal and conveniently define the following `even-odd' basis where 
\begin{align}
G_{even}=\frac{G^+(\omega,r,m)+G^+(-\omega,r,-m)}{2} \qquad G_{odd}=\frac{G^+(\omega,r,m)-G^+(-\omega,r,-m)}{2}\ .
\label{gevenodd}
\end{align}
In terms of this basis the full scalar solution can be written as 
\begin{align}
  \Phi&= G_{even}\phi_{even}+G_{odd}\phi_{odd}
\end{align}
with
\begin{subequations}
\begin{align}
  \phi_{even}&= -\phi_F+e^{\beta \left( \omega \left( 1-\chi  \right)-m\mu \left( 1-\tilde{\Theta } \right) \right)}\phi_P\\
  &= \left[ \phi_a-\left( \frac{e^{\beta \left( \omega (1-\chi )-m\mu (1-\tilde{\Theta }) \right)}-1}{e^{\beta \left( \omega -m\mu \tilde{\Theta } \right)}-1}-\frac{1}{2} \right)\phi_d \right]\\
  &= \left( 1-e^{-\beta \left( \omega \chi -m\mu \tilde{\Theta } \right)} \right)\left( f_{\omega ,m}+1 \right)\phi_R+\left( e^{\beta \left( \omega (1-\chi )-m\mu (1-\tilde{\Theta }) \right)}-1 \right)f_{\omega ,m}\phi_L\ ,\\
 \phi_{odd}&=  -\phi_F-e^{\beta \left( \omega \left( 1-\chi  \right)-m\mu \left( 1-\tilde{\Theta } \right) \right)}\phi_P\\
  &= \left[ \phi_a+\left( \frac{e^{\beta \left( \omega (1-\chi )-m\mu (1-\tilde{\Theta }) \right)}+1}{e^{\beta \left( \omega -m\mu \tilde{\Theta } \right)}-1}+\frac{1}{2} \right)\phi_d \right]\\
  &= \left( 1+e^{-\beta \left( \omega \chi -m\mu \tilde{\Theta } \right)} \right)\left( f_{\omega ,m}+1 \right)\phi_R-\left( e^{\beta \left( \omega (1-\chi )-m\mu (1-\tilde{\Theta }) \right)}+1 \right)f_{\omega ,m}\phi_L\ .
  \label{phiodd}
\end{align}
\end{subequations}
and we have normalised $\Theta$ in terms of $\tilde{\Theta}$ where $\tilde{\Theta}=\frac{\Theta}{\mu}$. The full solution becomes non-analytic due to the presence of $\chi$ and $\Theta$ in the outgoing solution. 
The full solution upto second order in derivatives can be written as 
\begin{align}
  \Phi
  &= \left[ 1+\left( \frac{\beta \omega }{2} \right)^2G^+_{\omega^2}+\left( \frac{\beta m\mu }{2} \right)^2 G^+_{m^2}+\left( \frac{\beta \omega }{2} \right)\left( \frac{\beta m\mu }{2} \right)G^+_{\omega m} \right]\phi_{even}\nonumber\\
  &\quad+\left[ G^+_{\omega }\left(\frac{\beta\omega}{2}\right)+G^+_{m}\frac{\beta m\mu }{2} \right]\phi_{odd}+{\text{higher order terms}}\ .
  \label{Phiderivativeexpansion}
\end{align}
Now we obtain an important relation using equations \eqref{eq:chi}, \eqref{eq:theta}, \eqref{eqq1} and \eqref{eqq2} that we explicitly use in the computation of quartic influence functional. The relation is given by
\begin{align}
G_{\omega}^++\chi &=-G_m^++\tilde{\Theta}\ .
\label{imp}
\end{align}
Similar relation holds true even for the massive case. See appendix \ref{sec:derivative expansion} for the derivation of the relation in massive case.\\
{\bf{Quartic Influence-functional:}}\\
Equipped with the derivative expanded full solution given in equation \eqref{Phiderivativeexpansion}, we now proceed to compute the quartic influence functional by substituting the solution in equation \eqref{squartic0}. Substituting the derivative expansion for $\phi_{even}$ the quartic influence functional in frequency domain becomes
\begin{align}
\begin{split}
S_{quartic}&= \int \prod_{i=1}^4\frac{d\omega }{2\pi }\left( \frac{1}{2\pi } \right)^{4}\sum_{m_1,m_2,m_3,m_4}\left\{ \mathcal{G}_{aaaa}\phi_a\left( \omega _1,m_1 \right)\phi_a\left( \omega _2,m_2 \right)\phi_{a}\left( \omega _3,m_3 \right)\phi_{a}\left( \omega _4,m_4 \right)\right.\\
&\left.\qquad \qquad \qquad \qquad \qquad \mathcal{G}_{aaad}\ \phi_{a}\left( \omega _1,m_1 \right)\ \phi_{a}\left( \omega _2,m_2 \right)\phi_{a}\left( \omega _3,m_3 \right)\phi_{d}\left( \omega _4,m_4 \right)\right.\\
&\left.\qquad \qquad \qquad \qquad \qquad+\mathcal{G}_{aadd}\ \phi_{a}\left( \omega _1,m_1 \right)\ \phi_{a}\left( \omega _2,m_2 \right)\phi_{d}\left( \omega _3,m_3 \right)\phi_{d}\left( \omega _4,m_4 \right)\right.\\
&\left.\qquad \qquad \qquad \qquad \qquad \mathcal{G}_{addd}\  \phi_{a}\left( \omega _1,m_1 \right)\phi_{d}\left( \omega _2,m_2 \right)\phi_{d}\left( \omega _3,m_3 \right)\phi_{d}\left( \omega _4,m_4 \right)\right.\\
&\left.\qquad \qquad \qquad \qquad \qquad+\mathcal{G}_{dddd}\ \phi_{d}\left( \omega _1,m_1 \right)\phi_{d}\left( \omega _2,m_2 \right)\phi_{d}\left( \omega _3,m_3 \right)\phi_{d}\left( \omega _4,m_4 \right) \right\}\ .
\end{split}
  \label{quarticinf}
\end{align}
The coefficient functions that enter in front of various terms in the quartic on-shell action are given by the following radial integrals over complexified radial coordinate. Note that we have performed the $v$ and $\phi'$ integrals in the following expressions that result in total energy and momentum conserving delta functions. The coefficient functions till first order in derivative expansion are
\begin{subequations}
\begin{align}
 \mathcal{G}_{aaaa}&=\frac{\lambda }{4!}(2\pi )\delta \left( \sum_i \omega _i \right)(2\pi )\delta _{\sum_i m_i}\oint d\chi \sqrt{-g}= 0\\
  \mathcal{G}_{aaad}&= \frac{\lambda }{3!}(2\pi )\delta \left( \sum_i \omega _i \right)(2\pi )\delta _{\sum_i m_i}\oint d\chi \sqrt{-g}\left\{ X-\beta \frac{\left( \omega _4-\mu m_4 \right)}{2}\left(X^2-\frac{1}{4}\right) \right\}\\
  \mathcal{G}_{aadd}&= \frac{\lambda }{4}(2\pi )\delta \left( \sum_i \omega _i \right)(2\pi )\delta _{\sum_i m_i}\oint d\chi \sqrt{-g}\left\{ X^2-\beta\left(\omega_4-\mu m_4\right) \left(X^3-\frac{X}{4}\right)  \right\}\\
  \begin{split}
    \mathcal{G}_{addd}&=\frac{\lambda }{3!}(2\pi )\delta \left( \sum_i \omega _i \right)(2\pi )\delta _{\sum_i m_i}\oint d\chi \sqrt{-g}\left\{X^3+\frac{\beta(\omega_4-\mu m_4)}{2}\left(X^4-\frac{X^2}{4}\right)\right\}
    \end{split}\\
  \begin{split}
  \mathcal{G}_{dddd}&= \frac{\lambda }{4!}(2\pi )\delta \left( \sum_i \omega _i \right)(2\pi )\delta _{\sum_i m_i}\oint d\chi \sqrt{-g} X^4 
  \end{split}
  \label{spectralfns}
\end{align}
\end{subequations}
where $X$ inside the integrals is independent of $\omega$ and $m$ and is defined as
\begin{align}
X= G_{\omega }^++\chi -\frac{1}{2}\ .
\label{X}
\end{align}
We have used the relation \eqref{imp} in obtaining the above integral expressions for the coefficient functions. $\mathcal{G}_{aaaa}=0$ since the radial integral is completely analytic. This is a consequence of microscopic unitarity or equivalently a consequence of Schwinger-Keldysh collapse rule that states that under collapse of two consecutive legs of the SK contour, the effective action vanishes.\\
{\bf{Evaluation of quartic integrals:}}\\
To evaluate the above integrals we note that the complexified contour integral can be broken into two parts as given below
\begin{align}
\oint d\chi=\int_{r_c}^{r_+} dr \frac{dr}{d\chi}+\int_{r_+}^{r_c}dr \frac{dr}{d\chi}
\end{align} 
where the integrand picks up a monodromy traversing across the branch point at $r=r_+$. This monodromy is correctly incorporated by the following substitution in the integrand as one goes across the branch point
\begin{subequations}
\begin{align}
\chi-\frac{1}{2}&\rightarrow \chi+\frac{1}{2}\\
\tilde{\Theta}-\frac{1}{2}&\rightarrow \tilde{\Theta}+\frac{1}{2}\ .
\end{align}
\end{subequations}
Since there are no other poles of the integrand at $r=r_+$, the above substitution in the integrand gives the value of the radial integration. 
\par
We evaluate the integrals in the slow rotation limit as a series in $\mu$. We will only keep terms that are linear order in $\mu$. The `bare' values of the integrals till linear order in small rotation and linear order in derivative expansion are the following 
\begin{subequations}
\begin{align}
\mathcal{G}_{aaaa}&=0\\
\mathcal{G}_{aaad}&= \frac{\lambda }{3!}(2\pi )\delta \left( \sum_i \omega _i \right)(2\pi )\delta _{\sum_i m_i}\left\{\frac{r_c^2}{2}-\frac{r_+^2}{2}+\frac{\beta}{2\pi}\left(\omega_4-\mu m_4\right)\log\left(\frac{r_c}{r_+}\right)\right\}+\mathcal{O}(\mu^2)\\
\mathcal{G}_{aadd}
&=\frac{\lambda }{4}(2\pi )\delta \left( \sum_i \omega _i \right)(2\pi )\delta _{\sum_i m_i}\left\{\frac{i}{\pi}r_+^2\log\left(\frac{r_c}{r_+}\right)-\frac{\beta}{8}r_+^2\left(\omega_4-\mu m_4\right)\right\}+\mathcal{O}(\mu^2)\\
\begin{split}
\mathcal{G}_{addd}
&=\frac{\lambda }{3!}(2\pi )\delta \left( \sum_i \omega _i \right)(2\pi )\delta _{\sum_i m_i}\left\{\frac{r_c^2}{8}-\frac{r_+^2}{4}+\frac{i\beta}{8\pi}\left(\omega_4-\mu m_4\right)\left(\frac{6}{\pi^2}r_+^2\zeta[3]-r_+^2\log\left(\frac{r_c}{r_+}\right)\right)\right\}\nonumber\\
&\qquad \qquad \qquad \qquad  \qquad \qquad \qquad \qquad \qquad \qquad  \qquad \qquad \qquad \qquad \qquad \   +\mathcal{O}(\mu^2)
\end{split}\\
\mathcal{G}_{dddd}
&=\frac{\lambda }{4!}(2\pi )\delta \left( \sum_i \omega _i \right)(2\pi )\delta _{\sum_i m_i}\left\{-ir_+^2\frac{3\zeta[3]}{2\pi^3}+i\frac{r_+^2}{2\pi}\log\left(\frac{r_c}{r_+}\right)\right\}+\mathcal{O}(\mu^2)\ .
\end{align}
\end{subequations}
The values of the integrals are divergent. Hence we need to introduce counter terms to kill these divergences and get finite answers. We do a minimal subtraction where we  only kill the divergent pieces and keep the finite part as it is. 
 From the above values of quartic influence functional, one clearly sees divergences that are both of polynomial type and logarithmic type in $r_c$. The polynomial divergences can be removed by adding local counter terms to the theory. However, the logarithmic divergences can not be removed by local counter terms since the argument of the logarithm has dependence on both $r_c$ and $r_+$. It turns out that, to renormalise the influence phase in the massless case, one has to do an additional source renormalisation \cite{Jana:2020vyx} that removes the logarithmic divergences. In the massive case one only encounters polynomial divergences that can be removed by local counter terms, hence source renormalisation is not needed. 
 \par
 In the following we discuss the divergence structure of the massless integrals only till linear order in small $\mu$ and linear order in derivative expansion. We will then find the counter terms (till linear order in small $\mu$ and linear order in derivative expansion), that cancel the divergences of the bare integrals. The details of the source renormalisation is discussed in section \ref{sec:renorm} when we consider arbitrary higher order terms in the effective action. 
Note that the divergence of the influence phase comes from integrals of the form \cite{Jana:2020vyx}
\begin{align}
F_k&=\oint d\chi\sqrt{-g}\left(X+G_{\omega}-\frac{1}{2}\right)^k\ .
\end{align}
The divergences of $F_k$ for $k=1,2,3,4$ till linear order in $\mu$ are of the forms
\begin{subequations} 
\begin{align}
F_1 &\sim \frac{r_c^2}{2}\\
F_2 &\sim \frac{ir_+^2}{\pi}\ln\left(\frac{r_c}{r_+}\right)\\
F_3 &\sim \frac{r_c^2}{8}\\
F_4 &\sim \frac{ir_+^2}{2\pi}\ln\left(\frac{r_c}{r_+}\right)\ .
\end{align}
\end{subequations}
We add the following counter terms (till linear order in $\mu$) to regularise the integrals
\begin{subequations}
\begin{align}
\delta_1&=-\frac{r_c^2}{2}\\ 
\delta_2&=-\frac{ir_+^2}{\pi}\ln\left(\frac{r_c}{r_+}\right)\\
\delta_3&=-\frac{r_c^2}{8}\\
\delta_4&=-
\frac{ir_+^2}{2\pi}\ln\left(\frac{r_c}{r_+}\right)\ .
\end{align}
\end{subequations}
Here $\delta_1$ and $\delta_2$ are local counter terms. However $\delta_2$ and $\delta_4$ are not. The origin of $\delta_2$ and $\delta_4$ lies within the source renormalisation scheme that we discuss in section \ref{sec:renorm}. 
\subsection{General $n$-th order effective theory}
\label{subn}
In this section we extend our study of effective theory to general n-point self interaction of the scalar field.  We will compute the $n$-th order influence functional following the same recipe discussed in the previous section. The coefficient functions appearing in front of various terms of the effective action are expressed in terms of integrals over the complexified radial contour. We will now consider the more complicated massive scalar case for our analysis since we will express the coefficient functions in the effective theory only as radial integrals. The action of the scalar field is given by
\begin{align}
  S&= -\int d\chi dv d\phi \sqrt{-g}\left\{ \frac{1}{2}\partial_{A}\Phi g^{A B}\partial_{B}\Phi  +M^2\Phi^2+\frac{\lambda _n}{n!} \Phi ^n\right\}\ .
\end{align}
The influence phase is again given by the on-shell action. 
Like quartic on-shell action, the $n$-point on-shell action becomes
\begin{align}
  S_{n-point}&= -\int d\chi dvd\phi \sqrt{-g} \frac{\lambda _n}{n!}\Phi _0^n
  \label{n point1}
\end{align}
where $\Phi _0$ satisfies the homogeneous wave equation
\begin{align}
  \frac{1}{\sqrt{-g}}\partial_{A}\left( \sqrt{-g}g^{A B} \partial_{B}\Phi _0 \right)-M^2 \Phi _0=0\ .
\end{align}
The derivation of this equation is exactly same as given in the previous section. We will write down the influence phase in position space upto first order in derivatives. Let us start with an expression for the $\Phi _0$ in position space written in average-difference/Keldysh basis
\begin{align}
\begin{split}
  \Phi _0 &= G_0^+\left( 1+\frac{i\beta }{2}\left( \tilde{G}_{\omega }^+\partial_{v}-\mu \tilde{G}_m^+\partial_{\phi' } \right) \right)\left\{ \phi _a^b+\frac{i\beta }{8}\left( \partial_{v}-\mu \partial_{\phi' } \right)\phi _d^b+\phi _d^b \left( \chi +\tilde{G}_{\omega }^+-\frac{1}{2} \right)\right.\\
  &\left.\hspace{5cm}-\frac{i\beta }{2}\left( \partial_{v}-\mu \partial_{\phi' } \right)\phi _d^b\left( \chi +\tilde{G}_{\omega }^+-\frac{1}{2} \right)^2 \right\}+{\text{higher order terms}}\ .
  \end{split}
  \label{phi0}
\end{align}
The superscript `b' in the above expression denotes that the sources at infinity are bare sources. 
We substitute equation \eqref{phi0} in equation \eqref{n point1} to obtain
\begin{align}
\begin{split}
  \frac{\Phi _0^n}{n!}&= (G_0^+)^n\sum_{k=0}^{n}\frac{\left( \phi _a^b +\frac{i\beta }{8}\left( \partial_{v}-\mu \partial_{\phi' } \right)\phi _d^b\right)^{n-k}}{(n-k)!}\frac{\left( \phi _d^b X-\frac{i\beta }{2}\left( \partial_{v}-\mu \partial_{\phi'} \right)\phi _d^b X^2 \right)^k}{k!}\\
  &\quad + \partial_{t}\left\{\frac{i\beta }{2(n!)}(G_0^+)^n\tilde{G}_{\omega }^+\left( \phi _a^b+\phi _d^b X \right)^n  \right\}-\mu \partial_{\phi' }\left\{ \frac{i\beta }{2(n!)}(G_0^+)^n\tilde{G}_m^+\left( \phi _a^b+\phi _d^b X \right) \right\}
  \end{split}
  \label{phi01}
\end{align}
where 
\begin{align}
X=\tilde{G}_{\omega}^+ +\chi-\frac{1}{2}
\end{align} and $\tilde{G}_{\omega}^+=\frac{G_{\omega}^+}{G_0^+}$ for a massive scalar field. 
Since the last two terms in equation \eqref{phi01} are total derivatives, they do not contribute in the influence phase.
Expanding the terms inside the sum upto first order in derivatives we get
\begin{subequations}
\begin{align}
\begin{split}
  \frac{\Phi _0^n}{n!}&=(G_0^+)^n\sum_{k=0}^{n}\frac{\left( (\phi _a^b)^{n-k}+(n-k)(\phi _a^b)^{n-k-1}\frac{i\beta }{8}\left( \partial_{v}-\mu \partial_{\phi'} \right)\phi _d^b \right)}{(n-k)!}\\
   &\qquad \qquad \qquad \frac{\left( (\phi _d^b)^k X^k-k(\phi _d^b)^{k-1}X^{k+1}\frac{i\beta }{2}\left( \partial_{v}-\mu \partial_{\phi'} \right)\phi _d^b \right)}{k!}
  \end{split}\\
  \begin{split}
  &= (G_0^+)^n\left\{\sum_{k=0}^{n} \frac{(\phi _a^b)^{n-k}(\phi _d^b)^k}{(n-k)!k!}X^k+\sum_{k=0}^{n-1}\frac{(\phi _a^b)^{n-k-1}(\phi _d^b)^k}{(n-k-1)!k!}X^k\frac{i\beta }{8}\left( \partial_{v}-\mu \partial_{\phi'} \right)\phi _d^b\right.\\
  &\qquad \qquad \left.-\sum_{k=1}^{n}\frac{(\phi _a^b)^{n-k}(\phi _d^b)^{k-1}}{(n-k)!(k-1)!}X^{k+1}\frac{i\beta }{2}\left( \partial_{v}-\mu \partial_{\phi'} \right)\phi _d^b \right\}+{\text{higher order}}\ .
  \end{split}
   \label{inf1}
\end{align}
  \end{subequations}
Hence the bare influence functional becomes \footnote{Note that $v$ goes to $t$ at the boundary where $r^*$ goes to $0$, also $\phi^{\prime}$ goes to $\phi$ since $r^{\#}$ goes to $0$ at the boundary.}
  \begin{align}
\begin{split}
S^b&=-\lambda_n\int  dt d\phi \left\{\sum_{k=1}^{n}\frac{(\phi _a^b)^{n-k}(\phi _d^b)^k}{(n-k)!k!}\mathcal{F}^b_{n,k}+\sum_{k=1}^{n-1}\frac{(\phi _a^b)^{n-k}(\phi _d^b)^{k-1}}{(n-k)!(k-1)!}\mathcal{F}^b_{n,k-1}\frac{i\beta }{8}\left( \partial_{t}-\mu \partial_{\phi } \right)\phi _d^b\right.\\
  &\left.\hspace{2cm}-\sum_{k=1}^{n-1}\frac{(\phi _a^b)^{n-k}(\phi _d^b)^{k-1}}{(n-k)!(k-1)!}\mathcal{F}^b_{n,k+1}\frac{i\beta }{2}\left( \partial_{t}-\mu \partial_{\phi } \right)\phi _d^b \right\}
  \end{split}
\end{align} where
\begin{align}
\mathcal{F}^b_{n,k}=\int_{r_+}^{r_c} dr\: r\:(G_0^+)^n\left\{\left(\chi+\tilde{G}_{\omega}^++\frac{1}{2}\right)^k-\left(\chi+\tilde{G}_{\omega}^+-\frac{1}{2}\right)^k\right\}\ .
 \label{inf2}
\end{align}
We have also used the fact that $k=0$ term in the second sum in equation \eqref{inf1} vanishes after integrating over the holographic Schwinger Keldysh contour since the integrand is analytic. $k=n$ term in the third sum in this equation also does not contribute since that term is just a total derivative. The bare influence functional that is computed with the bare sources as boundary conditions is divergent. The divergence structure can be  understood for massive case by looking at the asymptotics of the solution as discussed in appendix \ref{sec:derivative expansion}. We cancel the divergences employing an appropriate renormalisation scheme as discussed in the next section.
\subsection{Renormalizing the $n$-th order effective theory}
\label{renormscheme}
\label{sec:renorm}
The function $\mathcal{F}^b_{n,k}$ in equation \eqref{inf2} is divergent. The divergence structures are discussed below. 
\textbf{Massive case:}
\\
The divergence coming from $X^{2k+1}$ is
\begin{align}
\mathcal{F}^b_{n,2k+1}=\mathcal{F}^r_{n,2k+1}+\frac{\Lambda_{\Delta}}{4^k}\ , \quad  \Lambda_{\Delta}=r_+^2\left(\frac{\tilde{\zeta}_c}{1-\mu^2}\right)^{(\nu-1)n}\frac{\tilde{\zeta}_c}{2(\nu-1)n+2}
\end{align}
where 
$\tilde{\mathcal{\zeta}}=\left(\frac{r}{r_+}\right)^2$ and rescaled conformal dimension  $\nu=\frac{\Delta}{2}$\ . Hence this is divergent for all $\nu$.
\\
For $X^{2k}$ i.e. for $\mathcal{F}^b_{n,2k}$ there is no divergence for $\nu<1$ that are relevant deformations in the CFT (see equation \eqref{div1} of appendix \ref{sec:derivative expansion}).\\
\textbf{Massless case:}\\
The divergence coming from $X^{2k+1}$ is
\begin{align}
\mathcal{F}^b_{n,2k+1}&=\mathcal{F}^r_{n,2k+1}+\frac{\delta}{4^k}
\end{align}
and for  $X^{2k}$ it is
\begin{align}
\mathcal{F}^b_{n,2k}&=\mathcal{F}^r_{n,2k}+k\frac{\eta}{4^{k-1}}
\end{align}
where the divergences are
\begin{align}
\delta=\frac{r_c^2}{2}\ , \qquad \eta=\frac{i(r_+^2-r_-^2)}{\pi}\ln\left(\frac{r_c}{r_+}\right)\ .
\end{align}
For the massive case we add local counter terms in the bare influence phase to cancel the polynomial divergences and obtain a finite influence phase. However the massless case is more tricky, since along with polynomial divergences we also get non local logarithmic divergences that can not be removed by adding local state independent counter terms in the bare influence phase.
 It turns out that if we renormalise the sources at the cutoff in a particular way such that the bare and renormalised sources agree when the cutoff is taken to infinity, the logarithmic divergences get cancelled and the renormalised effective action becomes finite \cite{Jana:2020vyx}. This source renormalisation prescription for the massless case is given by the following temperature dependent dressing of sources where the difference source mixes with the average source 
 \begin{align}
\phi^b_a=\phi^r_a-\frac{\eta}{2\delta}\phi^r_d\ , \qquad \phi^b_d=\phi^r_d+\frac{\eta}{2\delta}i\beta\left(\partial_t-\mu\partial_{\phi}\right)\phi^r_d\ .
\end{align}
Note that as we take $r_c\to \infty$,  $\lim_{r_c\to\infty}\frac{\eta}{\delta}=0$, i.e., the bare and the renormalised sources match at asymptotic infinity.
We consider the following counter term expressed in terms of state independent renormalised sources 
\begin{align}
S^{c.t}=\lambda_n\int dtd\phi \frac{\delta}{n!}\left[\left(\phi^r_a+\frac{\phi_d^r}{2}\right)^n-\left(\phi^r_a-\frac{\phi_d^r}{2}\right)^n\right]\ .
\end{align}
Then the statement is that when we add this to the bare influence phase the renormalised influence functional is free of nonlocal logarithmic divergences. The renormalised influence functional becomes
\begin{align}
S^r&=\lim_{r_c\rightarrow\infty}\left(S^b+S^{c.t}\right)\\
\begin{split}
&=-\lambda_n\int  dt d\phi \left\{\sum_{k=1}^{n} \frac{(\phi _a^r)^{n-k}(\phi _d^r)^k}{(n-k)!k!}\mathcal{F}^r_{n,k}+\sum_{k=1}^{n-1}\frac{(\phi _a^r)^{n-k}(\phi _d^r)^{k-1}}{(n-k)!(k-1)!}\mathcal{F}^r_{n,k-1}\frac{i\beta }{8}\left( \partial_{t}-\mu \partial_{\phi } \right)\phi _d^r\right.\\
  &\left.\hspace{2cm}-\sum_{k=1}^{n-1}\frac{(\phi _a^r)^{n-k}(\phi _d^r)^{k-1}}{(n-k)!(k-1)!}\mathcal{F}^r_{n,k+1}\frac{i\beta }{2}\left( \partial_{t}-\mu \partial_{\phi } \right)\phi _d^r \right\}\ .
  \end{split}
\end{align}

\section{Nonlinear Langevin dynamics in rotating plasma}
The effective theory of the open scalar field has a description in terms of a nonlinear Langevin equation with non-Gaussianities in the thermal fluctuations. In the following we discuss the dual stochastic description of the effective dynamics. 
\label{sec:Nonlinear Langevin dynamics}
\subsection{Deriving effective action via stochastic path integral}
We start with an ansatz for the stochastic dynamics governed by the non-linear Langevin equation in presence of rotation for the average field $\psi_a$ as follows (for non-rotating case see \cite{Jana:2020vyx})
\begin{align}
  \mathcal{E}[\psi _a,\eta ]&\equiv f\eta  \\
 \begin{split}  
 \mathcal{E} [\psi _a,\eta ]&\equiv -K \partial_{t}^2 \psi _a+D\partial_{\phi}^2\psi _a+F\partial_{t}\partial_{\phi}\psi _a+\gamma \left( \partial_{t}-\mu \partial_{\phi} \right)\psi _a+\bar{\mu}\psi _a\\
  &+\sum_{k=1}^{n-1}\left( \theta _k\frac{\eta ^{k-1}}{k!}\frac{\psi _a^{n-k}}{(n-k)!}+\bar{\theta }_{k}\frac{\eta ^{k-1}}{k!}\frac{\psi _a^{n-k-1}}{(n-k-1)!}\left( \partial_{t}-\mu \partial_{\phi} \right)\psi _a \right)\ .
  \end{split}
\end{align}
The stochastic variable $\eta $ is the thermal noise, $f$ is the strength of the noise, $Z$, $X$ and $Y$ introduce colors in the thermal noise, $\gamma$ is the damping constant, $\theta_k$ and $\bar{\theta}_k$ correspond to the thermal jitter on top of the damping constant and anharmonicities in the dynamics of the open scalar field. We consider the most general probability distribution upto two derivatives acting on the stochastic noise $\eta$. We take the distribution of the noise to be 
 \begin{align}
  \mathcal{P}[\eta ]&= \mathcal{N}e^{-\int dt d\phi \left( \frac{f}{2!}\eta ^2+\frac{Z}{2!} (\partial_{t}\eta)^2 +\frac{X}{2!} (\partial_{\phi}\eta)^2 +\frac{Y}{2!} \partial_{t}\eta \partial_{\phi}\eta +\frac{\theta _n}{n!}\eta ^n \right)}
  \label{noisedistr}
\end{align}
where $\mathcal{N}$ is a normalisation factor and $\theta _n$ is the non-Gaussianity in the noise. In the following we obtain an effective action from a stochastic path integral that we eventually compare with the influence functional obtained from holographic computations. This will determine the parameters of the non-linear Langevin equation in terms of the effective couplings computed from holography. To compute the effective action from stochastic path integral we will exploit Martin-Siggia-Rose (MSR) \cite{1973PhRvA...8..423M, 1978PhRvB..18..353D, 1976ZPhyB..23..377J} trick. For this we will have to introduce an auxiliary field $\psi_d$ and convert following identity
\begin{align}
  1&= \int D\psi _aD\eta \ \delta \left( \mathcal{E} [\psi _a,\eta ]-f\eta \right)\mathcal{P}[\eta ]
\end{align}
to the identity
\begin{align}
  1&= \int D\psi _aD\psi _dD\eta e^{-i\int d^2x \left( \mathcal{E} [\psi _a,\eta ]-f\eta  \right)\psi_d}\mathcal{P}[\eta ]\ .
\end{align}
Now we integrate over the noise $\eta$ to get an effective action in terms of $\psi _a$ and $\psi _d$. This can be done by shifting $\eta\rightarrow \eta+i\psi_d$ and taking $\eta\rightarrow 0$. This will give the leading order Schwinger-Keldysh effective action with non-zero rotation
\begin{align}
\begin{split}
S_{\psi }&= -\int dt d\phi\left[- K \psi _d\partial_{t}^2\psi _a+D\psi _d\partial_{\phi}^2\psi _a+F\psi _d\partial_{t}\partial_{\phi}\psi _a+\bar{\mu}\psi _a\psi _d+\gamma\:\psi _d\left( \partial_{t}-\mu \partial_{\phi} \right)\psi _a\right.\\
&\left.\hspace{3cm} -\frac{if}{2!}\psi _d^2+\frac{iZ}{2!}(\partial_{t}\psi _d)^2+\frac{iX}{2!}(\partial_{\phi }\psi _d)^2+\frac{iY}{2!}\partial_{t}\psi _d\partial_{\phi}\psi _d\right.\\
&\left.\hspace{2cm}-i\sum_{k=1}^n\theta _k\frac{(i\psi _d)^k}{k!}\frac{\psi _a^{n-k}}{(n-k)!}-i\sum_{k=1}^{n-1}\bar{\theta }_k\frac{(i\psi _d)^k}{k!}\frac{\psi _a^{n-k-1}}{(n-k-1)!}\left( \partial_{t}-\mu \partial_{\phi} \right)\psi _a\right]\ .
\end{split}
\label{Spsi}
\end{align}
\subsection{Comparing with holographic result}
In this section we will compare the effective action derived from stochastic path integral to the effective action computed using holography. This way we evaluate the parameters in the non-linear Langevin equation in terms of the effective couplings computed from holography. 
The way to do this is by uplifting the sources $\phi _a$ and $\phi _d$ that appear in the influence functional to stochastic fields that we denote by $\psi _a$ and $\psi _d$. Let us begin by writing down the exact quadratic influence functional 
\begin{align}
\begin{split}
  S_{quadratic}&= \mathcal{N}_2\int dt d\phi \left\{ i\frac{\psi _d^2}{\beta }+i\mathfrak{h}_{1,1}\partial_{t}\psi _d\partial_{\phi}\psi _d+i\mathfrak{h}_{0,2}\left( \partial_{\phi}\psi _d \right)^2+i\mathfrak{h}_{2,0}\left( \partial_{t}\psi _d \right)^2 \right.\\
  &\left.+\mathfrak{g}_{0,0}\psi _a\psi _d-\left( \partial_{t}\psi _a-\mu \partial_{\phi}\psi _a \right)\psi _d+\mathfrak{g}_{0,2}\partial_{\phi}\psi _a\partial_{\phi}\psi _d+\mathfrak{g}_{2,0}\partial_{t}\psi _a\partial_{t}\psi _d+\mathfrak{g}_{1,1}\partial_{t}\psi _a\partial_{\phi}\psi _d\right\}
  \end{split}
  \label{stochquadratic}
\end{align}
where
\begin{align}
\mathcal{N}_2&=-\frac{(r_+^2-r_-^2)^{\Delta-2}\Gamma\left(\frac{\Delta}{2}\right)^4 r_+}{\Gamma(\Delta-1)^2}\ .
\end{align}
The exact coefficient functions computed from holography are 
\begin{subequations}
\begin{align}
  \mathfrak{h}_{0,2}&= \frac{\beta\left[\pi^2\mu^2-(1+\mu^2)\mathcal{\psi}^1\left(\frac{\Delta}{2}\right)\right]}{8\pi^2}\\
  \mathfrak{h}_{2,0}&=\frac{\beta\left[\pi^2-(1+\mu^2)\mathcal{\psi}^1\left(\frac{\Delta}{2}\right)\right]}{8\pi^2} \\
 \mathfrak{h}_{1,1}&=\frac{\beta\mu\left[\pi^2-2\mathcal{\psi}^1\left(\frac{\Delta}{2}\right)\right]}{4\pi^2}\\
  \mathfrak{g}_{0,0}&= \frac{2\tan\left(\frac{\pi\Delta}{2}\right)}{\beta}\\
  \mathfrak{g}_{0,2}&= -\frac{\beta\left[\pi^2(-1+\mu^2\cos(\pi\Delta))\csc(\pi\Delta)+(1+\mu^2)\mathcal{\psi}^1\left(\frac{\Delta}{2}\right)\tan\left(\frac{\pi\Delta}{2}\right)\right]}{4\pi^2}\\
  \mathfrak{g}_{2,0}&=  -\frac{\beta\left[\pi^2(-\mu^2+\cos(\pi\Delta))\csc(\pi\Delta)+(1+\mu^2)\mathcal{\psi}^1\left(\frac{\Delta}{2}\right)\tan\left(\frac{\pi\Delta}{2}\right)\right]}{4\pi^2}\\
  \mathfrak{g}_{1,1}&= \frac{\beta\mu\left[\pi^2-2\mathcal{\psi}^1\left(\frac{\Delta}{2}\right)\right]\tan\left(\frac{\pi\Delta}{2}\right) }{2\pi^2}
\end{align}
\end{subequations}
where $\mathcal{\psi}^{1}$ is the Polygamma function of order one. 
\par
The $n$-th order influence functional computed from holography is given by
\begin{align}
\begin{split}
S^{r}_{\Phi^n}&= -\lambda _n \int dt d\phi \left\{ \sum_{k=1}^{n}\frac{\left( \psi _a^r \right)^{n-k}\left( \psi _d^r \right)^k}{(n-k)!k!}\mathcal{F}_{n,k}^r\right.\\
&\left.\qquad-\frac{i\beta }{2}\sum_{k=1}^{n-1}\frac{\left( \psi _a^r \right)^{n-k}\left( \psi _d^r \right)^{k-1}}{(n-k)!(k-1)!}\left( \partial_{t}-\mu \partial_{\phi} \right)\psi _d^r\: \left(\mathcal{F}_{n,k+1}^r-\frac{1}{4}\mathcal{F}_{n,k-1}^r\right)\right\}\ .
\end{split}
  \label{Phi^n}
\end{align}
Combining both the quadratic and $n$-th order influence functional we finally have
\begin{align}
\begin{split}
  S&= \int dt d\phi \left[\mathcal{N}_2\left\{ i\frac{\psi _d^2}{\beta }+i\mathfrak{h}_{1,1}\partial_{t}\psi _d\partial_{\phi}\psi _d+i\mathfrak{h}_{0,2}\left( \partial_{\phi}\psi _d \right)^2+i\mathfrak{h}_{2,0}\left( \partial_{t}\psi _d \right)^2 +\mathfrak{g}_{0,0}\psi _a\psi _d\right.\right.\\
  &\hspace{1cm}\left.\left.-\left( \partial_{t}\psi _a-\mu \partial_{\phi}\psi _a \right)\psi _d+\mathfrak{g}_{0,2}\partial_{\phi}\psi _a\partial_{\phi}\psi _d+\mathfrak{g}_{2,0}\partial_{t}\psi _a\partial_{t}\psi _d+\mathfrak{g}_{1,1}\partial_{t}\psi _a\partial_{\phi}\psi _d\right\}\right.\\
  &\left.-\lambda_n\sum_{k=1}^{n}\frac{\left( \psi _a^r \right)^{n-k}\left( \psi _d^r \right)^k}{(n-k)!k!}\mathcal{F}_{n,k}^r+\lambda_n\frac{i\beta }{2}\sum_{k=1}^{n-1}\frac{\left( \psi _a^r \right)^{n-k}\left( \psi _d^r \right)^{k-1}}{(n-k)!(k-1)!}\left( \partial_{t}-\mu \partial_{\phi } \right)\psi _d^r\: \left(\mathcal{F}_{n,k+1}^r-\frac{1}{4}\mathcal{F}_{n,k-1}^r\right)\right].
  \end{split}
  \label{stoch}
\end{align}
Now comparing equation \eqref{stoch} with equation \eqref{Spsi} we obtain the parameters of the non-linear Langevin equation in terms of the effective couplings computed using holography. The values of the parameters of the linear Langevin equation are
\begin{subequations}
\begin{align}
K&=-\mathcal{N}_2 \mathfrak{g}_{2,0}\qquad
D=\mathcal{N}_2 \mathfrak{g}_{0,2}\\
F&=\mathcal{N}_2 \mathfrak{g}_{1,1}\qquad\quad
\bar{\mu}=-\mathcal{N}_2 \mathfrak{g}_{0,0}\\
\gamma&=\mathcal{N}_2\qquad\qquad\quad
f=\frac{2\mathcal{N}_2}{\beta}\\
Z&=-2\mathcal{N}_2\mathfrak{h}_{2,0}\qquad
X=-2\mathcal{N}_2\mathfrak{h}_{0,2}\\
Y&=-2\mathcal{N}_2\mathfrak{h}_{1,1}\ .
\end{align}
\end{subequations}
Comparing the values of the parameters we see that the damping constant is related to the fluctuation as below
\begin{align}
f&=\frac{2\gamma}{\beta}\ .
\end{align}
This is the fluctuation dissipation relation between noise and dissipation in the linear Langevin equation with rotation. 
\par
The extension to non-linear Langevin equation has additional parameters like jitter in the damping, anharmonicity in the dynamics and non-Gaussianity in the thermal noise. Those additional parameters are $\theta_k$ and $\bar{\theta}_k$. From holographic computations we find 
\begin{subequations}
\begin{align}
\theta_k&=\frac{\lambda_n}{i^{k+1}}\mathcal{F}_{n,k}^r\\
\bar{\theta}_k&=\frac{\beta\lambda_n}{2i^{k+2}}\left(\mathcal{F}_{n,k+1}^r-\frac{1}{4}\mathcal{F}_{n,k-1}^r\right)
\end{align}
\end{subequations} 
where $\mathcal{F}_{n,k}^r$ are radial integrals that enter the effective couplings. 
From these we get the generalised fluctuation dissipation relation given by
\begin{align}
\frac{2}{\beta}\bar{\theta}_k&=\left(\theta_{k+1}+\frac{1}{4}\theta_{k-1}\right)\ .
\label{nonlinear FDR}
\end{align}
We have determined $\mathcal{F}_{n,k}^r$ for the quartic order effective theory by performing the radial integrals analytically in the slow rotation limit till linear order in rotation. 
In this case we find that generalised fluctuation dissipation relation given in equation \eqref{nonlinear FDR} holds true for linear order in slow rotation. 
\section{Conclusion and Discussion}
\label{sec:Conclusion}
In this paper, we consider a natural extension of the gravitational Schwinger-Keldysh path integral prescription of \cite{Glorioso:2018mmw} to rotating BTZ black holes. The gravitational space-time asymptotes to the real-time Schwinger-Keldysh contour of the dual rotating CFT at a given initial state with finite temperature and chemical potential due to angular momentum. We study a probe scalar field in the rotating BTZ geometry and obtain the ingoing quasi normal modes and outgoing (time-reversed) Hawking radiation in section \ref{sec:Probe scalar}. By computing the on-shell action and integrating over the complexified radial coordinate we construct the effective theory of an open scalar field that is coupled to a two-dimensional rotating thermal CFT plasma at the boundary in section \ref{sec:Influence functional}. The quadratic influence functionals computed in section \ref{quad1} match with the results of \cite{Son:2002sd} under appropriate coordinate transformations and redefinition of parameters. The quadratic effective theory has a dual stochastic description in terms of a linear Langevin equation in presence of rotation. The noise and dissipation terms in the Langevin equation are related by the fluctuation-dissipation relation with chemical potential.\par 
We determine the higher order terms in the effective theory by derivative expanding the free scalar solution in low frequency and angular momentum. The coefficient functions that appear in front of the higher order terms are the effective couplings evaluated from holography. We determine these coefficient functions of higher order terms in the form of integrals over the bulk contour in section \ref{subn}. In the limit when the CFT plasma has low angular momentum i.e. the dual BTZ black hole is slowly rotating, we compute the quartic order integrals analytically till first order in slow rotation (similar thing can be done for cubic effective theory also) in section \ref{sub4}. We find the correct renormalisation scheme to remove the divergences appearing in the values of the bare integrals in section \ref{renormscheme}. 
The higher order effective theory has a description in terms of a non-linear Langevin dynamics with non-Gaussianity in the thermal fluctuations as obtained in section \ref{sec:Nonlinear Langevin dynamics}. We write down the generalisation of the non-linear fluctuation dissipation relation in presence of chemical potential (see \cite{Jana:2020vyx} for the non-rotating case). 
\par
It will be useful to do the integrals that appear as the coefficient functions in the effective theory numerically. Employing a systematic holographic renormalisation scheme as given in \cite{Skenderis:2008dg}, will provide a better handle on the divergences coming from evaluating the integrals in our analysis. As a followup to our work, it will be interesting 
to generalise the prescription to higher-dimensional rotating black holes. Using derivative expansion in low frequency and angular momenta one can obtain the effective theory holographically. It will also be interesting in higher dimension to add a Chern-Simons term in the bulk since such a term in the bulk will lead to an anomaly on the even dimensional rotating boundary CFT.  Chern-Simons term in the presence of rotation is expected to give rise to an anomaly induced transport. 
\par
It will also be quite interesting to extend the holographic path integral contour to near-extremal black holes and make connection with the derivative expansion scheme 
given in \cite{moitra2020near} for near-extremal black branes. 
\section*{Acknowledgements}
We are grateful to R. Loganayagam for suggesting the problem to us and having several useful discussions during the course of this project. We would like to thank Soumyadeep Chaudhury and Amitabh Virmani for reading the draft and sending their comments. We would like to thank Mukund Rangamani and Amitabh Virmani for pointing out relevant references. We also thank Chandan Jana, Rohan Poojary and Akhil Sivakumar for discussing with us.
 
\appendix
\section{Quadratic Influence functionals in RA basis}
In the following we compute the quadratic influence functional in retarded-advanced basis. 
\label{sec:integrals1}
\begin{subequations}
\begin{align}
  S^{on-shell}&= -\int dv d\varphi ' \left[ \frac{r}{2}\Phi \partial_{v}\Phi -\frac{ir}{\beta }\Phi \partial_{\chi }\Phi +\frac{n_{\phi} r}{2}\Phi \partial_{\varphi '}\Phi  \right]\bigg{\rvert}^{\chi =1,  \Theta=\mu_+}_{\chi =0,  \Theta=0}\\
 \begin{split}
  &= -\int dv d\varphi '\left[ \int \frac{d\omega_1 d\omega _2}{(2\pi)^2 }\frac{1}{(2\pi)^2 }\sum_{m_1,m_2}\left( -G^+\left( \omega _1,r,m_1 \right)\phi_F\left( \omega _1,m_1 \right)\right.\right.\\
  &\left.\left.+G^+\left( -\omega _1,r,-m_1 \right)\phi_P\left( \omega _1,m_1 \right)e^{\beta \left( \omega _1\left( 1-\chi  \right)-m_1\left( \mu_+ -\Theta  \right) \right)} \right)e^{-i(\omega _1+\omega_2)v+i(m_{1}+m_2)\varphi '}\right.\\
&\left. \left(-\phi_F\left( \omega _2,m_2 \right)\left( \frac{r}{2}(-i\omega _2)G^+\left( \omega _2,r,m_2 \right)-\frac{ir}{\beta }\partial_{\chi }G^+\left( \omega _2,r,m_2 \right)+\frac{n_{\phi} r}{2}(im_2)G^+\left( \omega _2,r,m_2 \right) \right)\right.\right.\\
&\left.\left. +\phi_P\left( \omega _2,m_2 \right)e^{\beta \left( \omega _2\left( 1-\chi  \right)-m_2\left( \mu_+ -\Theta  \right) \right)}\left( \frac{r}{2}(i\omega _2)G^+\left( -\omega _2,r,-m_2 \right)-\frac{ir}{\beta }\partial_{\chi }G^+\left( -\omega _2,r,-m_2 \right)\right.\right.\right.\\
&\left.\left.\left.-\frac{n_{\phi} r}{2}(im_2)G^+\left( -\omega _2,r,-m_2 \right) \right)\right)\right]\bigg{\rvert}_{\chi =0,  \Theta=0}^{\chi =1,  \Theta=\mu_+}\ .
\end{split}
\end{align}
\end{subequations}
Integrating over $v$ and $\phi^{\prime}$ coordinates, we get energy and momenta conserving delta functions in the following
\begin{align}
\begin{split}
S^{on-shell}&=-\left[ \int \frac{d\omega_1 d\omega _2}{(2\pi)^2 }\frac{1}{(2\pi)^2 }\sum_{m_1,m_2}(2\pi)\delta(\omega_1+\omega_2)(2\pi)\delta_{m_1+m_2,0}\right.\\
&\left.\left\{-G^+\left( \omega _1,r,m_1 \right)\phi_F\left( \omega _1,m_1 \right)\phi_P\left( \omega _2,m_2 \right) e^{\beta \left( \omega _2\left( 1-\chi  \right)-m_2\left( \mu_+ -\Theta  \right) \right)}\right.\right.\\
&\left.\left.\left( \frac{r}{2}(i\omega _2)G^+\left( -\omega _2,r,-m_2 \right)-\frac{ir}{\beta }\partial_{\chi }G^+\left( -\omega _2,r,-m_2 \right)-\frac{n_{\phi} r}{2}(im_2)G^+\left( -\omega _2,r,-m_2 \right) \right)\right.\right.\\
&\left.\left. -G^+(-\omega_1,r,-m_1)\phi_P(\omega_1,m_1)e^{\beta(\omega_1(1-\chi)-m_1(\mu_+-\Theta))}\phi_F(\omega_2,m_2)\right.\right.\\
&\left.\left.\left( \frac{r}{2}(-i\omega _2)G^+\left( \omega _2,r,m_2 \right)-\frac{ir}{\beta }\partial_{\chi }G^+\left( \omega _2,r,m_2 \right)+\frac{n_{\phi} r}{2}(im_2)G^+\left( \omega _2,r,m_2 \right) \right)\right\}\right]\bigg\rvert^{\chi=1, \Theta=\mu_+}_{\chi=0,  \Theta=0}\ .
\end{split}
\end{align}
Since $\omega_1$ ,$\omega_2$ and $m_1$,$m_2$ are dummy variables, we will interchange them in the last four terms
\begin{align}
\begin{split}
S^{on-shell}&=-\int \frac{d\omega_1 d\omega _2}{(2\pi)^2 }\frac{1}{(2\pi)^2 }\sum_{m_1,m_2}(2\pi)\delta(\omega_1+\omega_2)(2\pi)\delta_{m_1+m_2,0}\\
&\left\{-G^+\left( \omega _1,r,m_1 \right)\phi_F\left( \omega _1,m_1 \right)\phi_P\left( \omega _2,m_2 \right)e^{\beta \left( \omega _2\left( 1-\chi  \right)-m_2\left( \mu_+ -\Theta  \right) \right)}\right.\\
&\left.\left( \frac{r}{2}(i\omega _2)G^+\left( -\omega _2,r,-m_2 \right)-\frac{ir}{\beta }\partial_{\chi }G^+\left( -\omega _2,r,-m_2 \right)-\frac{n_{\phi} r}{2}(im_2)G^+\left( -\omega _2,r,-m_2 \right) \right)\right.\\
&\left. -G^+(-\omega_2,r,-m_2)\phi_P(\omega_2,m_2)e^{\beta(\omega_2(1-\chi)-m_2(\mu_+-\Theta))}\phi_F(\omega_1,m_1)\right.\\
&\left.\left( \frac{r}{2}(-i\omega _1)G^+\left( \omega _1,r,m_1 \right)-\frac{ir}{\beta }\partial_{\chi }G^+\left( \omega _1,r,m_1 \right)+\frac{n_{\phi} r}{2}(im_1)G^+\left( \omega _1,r,m_1 \right) \right)\right\}\bigg{\rvert}_{\chi =0,  \Theta=0}^{\chi =1,  \Theta=\mu_+}\ .
\end{split}
\end{align}
Imposing the energy and momentum delta functions, gives the influence phase in the R-A basis
\begin{subequations}
\begin{align}
\begin{split}
S^{on-shell}&= -\int \frac{d\omega }{2\pi }\frac{1}{2\pi }\sum_m 2\phi_F\left( \omega ,m \right)\phi_P\left( -\omega ,-m \right)e^{-\beta \left( \omega \left( 1-\chi  \right)-m\left( \mu_+ -\Theta  \right) \right)}\\
&\qquad \frac{ir}{\beta }\left\{ G^+\left( \omega ,r,m \right)\left[ \partial_{\chi }G^+\left( \omega ,r,m \right) +\frac{\beta}{2}(\omega-n_{\phi}m)G^+(\omega,r,m)\right]\right\}\bigg{\rvert}_{\chi =0,  \Theta=0}^{\chi =1,  \Theta=\mu_+}
\end{split}\\
&= \int\frac{d\omega }{2\pi }\frac{1}{2\pi }\sum_m  \phi_F(\omega,m)\phi_P(-\omega,-m)\mathcal{G}_{FP}[\omega,m]\ .
\label{qdrI.F}
\end{align}
\end{subequations}
where
\begin{align}
\mathcal{G}_{FP}[\omega, m]&=-\frac{2ir}{\beta}G^{+}(\omega,r_c,m)\left[\partial_{\chi}G^+(\omega,r_c,m)+\frac{\beta}{2}(\omega-m n_{\phi})G^+(\omega,r_c,m)\right](1-e^{-\beta(\omega-\mu_+ m)})
\end{align} and other terms in the integral are not contributing because they are analytic. 
\section{Derivative expansion of massive scalar solution}
\label{sec:derivative expansion}
The massive scalar field equation is given by
\begin{align}
\begin{split}
 & \partial_{\chi }\left( r\partial_{\chi }G^+ \right)+\left( \frac{\beta \omega }{2} \right)\left( r\partial_{\chi }G^+ +\partial_{\chi }\left( rG^+ \right) \right)+\left( \frac{\beta m}{2} \right)^2\frac{f}{r}G^+\\
  & -\left( \frac{\beta m}{2} \right)2 n_{\phi}(r)r\partial_{\chi }G^+ -\frac{\beta m}{2}\partial_{\chi }\left( r n_{\phi}(r) \right)G^++\frac{\beta^2}{4}f r M^2 G^+ = 0\ .
  \end{split}
  \label{EOM2}
\end{align}
We substitute the derivative expansion given in equation \eqref{Gderivativeexpansion} in the above equation and solve the equation order by order in $\beta \omega$ and $\beta m \mu$.
Zeroth order solution is obtained by solving 
\begin{align}
\partial_r\left(r f \partial_r G_0^+\right)+r M^2 G_0^+=0\ .
\label{massiveG0}
\end{align}
Let us define a new variable and a parameter as follows
\begin{align}
\tilde{\mathcal{\zeta}}=\left(\frac{r}{r_+}\right)^2\ ,
\qquad \nu=\frac{\Delta}{2}\ .
\end{align}
In terms of this \eqref{massiveG0} can be rewritten as
\begin{align}
\partial_{\tilde{\mathcal{\zeta}}}\left((\tilde{\mathcal{\zeta}}-\mu^2)(\tilde{\mathcal{\zeta}}-1)\partial_{\tilde{\mathcal{\zeta}}}G_0^+\right)-\nu(\nu-1)G_0^+=0\ .
\end{align} 
The general solution to this 
\begin{align}
c_1 P_{-\nu}\left(\frac{2\tilde{\mathcal{\zeta}}-1-\mu^2}{1-\mu^2}\right)+c_2 Q_{-\nu}\left(\frac{2\tilde{\mathcal{\zeta}}-1-\mu^2}{1-\mu^2}\right)\ .
\end{align}
Now imposing the boundary condition $\frac{d G_0^+}{d\chi}=0$ at $r_+$ we remove the second solution. Normalizing the solution to one at the boundary fixes $c_1$. Finally we obtain
\begin{align}
G_0^+=\frac{P_{-\nu}\left(\frac{2\tilde{\mathcal{\zeta}}-1-\mu^2}{1-\mu^2}\right)}{P_{-\nu}\left(\frac{2\tilde{\mathcal{\zeta}}_c-1-\mu^2}{1-\mu^2}\right)}
\end{align}
where $\tilde{\zeta}_c$ denotes the value $\tilde{\mathcal{\zeta}}$ at the UV cut-off.
\par Now we move to the first order calculation:  $\tilde{G}_{\omega}^+=\frac{G_{\omega}^+}{G_0^+}$ satisfies
\begin{align}
\partial_{\tilde{\mathcal{\zeta}}}\left((\tilde{\mathcal{\zeta}}-\mu^2)(\tilde{\mathcal{\zeta}}-1)(G_0^+)^2\partial_{\tilde{\mathcal{\zeta}}}\tilde{G}_{\omega}^+\right)&=\frac{i(1-\mu^2)}{2\pi}\partial_{\tilde{\mathcal{\zeta}}}\left((G_0^+)^2\tilde{\mathcal{\zeta}}^{\frac{1}{2}}\right)\ .
\label{massive der0}
\end{align}
$\tilde{G}_m^+=\frac{G_{m}^+}{G_0^+}$ satisfies the following equation
\begin{align}
\mu \partial_{\tilde{\mathcal{\zeta}}}\left((\tilde{\mathcal{\zeta}}-\mu^2)(\tilde{\mathcal{\zeta}}-1)(G_0^+)^2\partial_{\tilde{\mathcal{\zeta}}}\tilde{G}_{m}^+\right)&=\frac{1-\mu^2}{2\pi i}\tilde{\mathcal{\zeta}}\left((G_0^+)^2\tilde{\mathcal{\zeta}}^{-\frac{1}{2}}\mu\right)\ .
\label{derim}
\end{align}
Note that the boundary condition satisfied by $\tilde{G}^+$ are 
\begin{align}
\frac{d}{d\chi }\tilde{G}^+_{\omega ^n,m^l}&=0 \qquad \tilde{G}^+_{\omega ^n,m^l}\rvert_{r_c}= 0\hspace{1cm} \forall n,l>0.
   \label{boundarycond}
\end{align}
Equation \eqref{massive der0} can be integrated to obtain $\tilde{G}_{\omega}^+$
\begin{subequations}
\begin{align}
\tilde{G}_{\omega}^+ &=\frac{i (1-\mu^2)}{2\pi}\int_{\tilde{\mathcal{\zeta}}_c}^{\tilde{\mathcal{\zeta}}}\frac{d\tilde{\mathcal{\zeta}}'}{(\tilde{\mathcal{\zeta}}'-\mu^2)(\tilde{\mathcal{\zeta}}'-1)(G_0^+[\tilde{\mathcal{\zeta}}'])^2}\int_1^{\tilde{\mathcal{\zeta}}'}\partial_{\tilde{\mathcal{\zeta}}''}\left((G_0^+[\tilde{\mathcal{\zeta}}''])^2(\tilde{\mathcal{\zeta}}'')^{\frac{1}{2}}\right)\\
&=\frac{i (1-\mu^2)}{2\pi}\int_{\tilde{\mathcal{\zeta}}_c}^{\tilde{\mathcal{\zeta}}}d\tilde{\mathcal{\zeta}}'\left\{\frac{\tilde{\mathcal{\zeta}}'^{\frac{1}{2}}}{(\tilde{\mathcal{\zeta}}'-\mu^2)(\tilde{\mathcal{\zeta}}'-1)}-\frac{(G_0^+[1])^2}{(\tilde{\mathcal{\zeta}}'-\mu^2)(\tilde{\mathcal{\zeta}}'-1)(G_0^+[\tilde{\mathcal{\zeta}}'])^2}\right\}\ .
\end{align}
\end{subequations}
Equation \eqref{derim} can be integrated to obtain $\tilde{G}_{m}^+$
\begin{subequations}
\begin{align}
\tilde{G}_{m}^+ &=\frac{ (1-\mu^2)}{2\pi i}\int_{\tilde{\mathcal{\zeta}}_c}^{\tilde{\mathcal{\zeta}}}\frac{d\tilde{\mathcal{\zeta}}'}{(\tilde{\mathcal{\zeta}}'-\mu^2)(\tilde{\mathcal{\zeta}}'-1)(G_0^+[\tilde{\mathcal{\zeta}}'])^2}\int_1^{\tilde{\mathcal{\zeta}}'}\partial_{\tilde{\mathcal{\zeta}}''}\left((G_0^+[\tilde{\mathcal{\zeta}}''])^2(\tilde{\mathcal{\zeta}}'')^{-\frac{1}{2}}\right)\\
&=\frac{ (1-\mu^2)}{2\pi i}\int_{\tilde{\mathcal{\zeta}}_c}^{\tilde{\mathcal{\zeta}}}d\tilde{\mathcal{\zeta}}'\left\{\frac{\tilde{\mathcal{\zeta}}'^{-\frac{1}{2}}}{(\tilde{\mathcal{\zeta}}'-\mu^2)(\tilde{\mathcal{\zeta}}'-1)}-\frac{(G_0^+[1])^2}{(\tilde{\mathcal{\zeta}}'-\mu^2)(\tilde{\mathcal{\zeta}}'-1)(G_0^+[\tilde{\mathcal{\zeta}}'])^2}\right\}\ .
\end{align}
\end{subequations}
In fact one can give a general recursive solution to $\tilde{G}_{\omega^l,m^k}$ in terms of lower order functions
\begin{align}
\begin{split}
\tilde{G}_{\omega^l,m^k}^+&=\int _{\tilde{\mathcal{\zeta}}_c}^{\tilde{\mathcal{\zeta}}
}\frac{1}{(\tilde{\mathcal{\zeta}}-\mu^2)(\tilde{\mathcal{\zeta}}-1)(G_0^+)^2}\int_1^{\tilde{\mathcal{\zeta}}'}d\tilde{\mathcal{\zeta}}'' \left\{\frac{1-\mu^2}{2\pi i}\left(2\tilde{\mathcal{\zeta}}''^{-\frac{1}{2}}\mu \partial_{\tilde{\mathcal{\zeta}}''}G^+_{\omega^l,m^{k-1}}+\partial_{\tilde{\mathcal{\zeta}}''}(\tilde{\mathcal{\zeta}}''^{-\frac{1}{2}}\mu)G^+_{\omega^l,m^{k-1}}\right.\right.\\
&\left.\left. \hspace{3cm}-2\tilde{\mathcal{\zeta}}''^{\frac{1}{2}}\partial_{\tilde{\mathcal{\zeta}}''}G^+_{\omega^{l-1},m^k}-G^+_{\omega^{l-1},m^k}\right)+\frac{(1-\mu^2)^2}{(2\pi)^2}\frac{G^+_{\omega^l,m^{k-2}}}{\tilde{\mathcal{\zeta}}''}\right\}G_0^+[\tilde{\mathcal{\zeta}}'']\ .
\end{split}
\end{align}
Here we assumed that $G^+_{\omega^l,m^k}$ with $l<0$ or $k<0$ is zero. With this one can check that the general formula reproduces the special cases.
We also note an important relation that can be used to compute the massive higher order influence functionals 
\begin{align}
\tilde{G}_{\omega}^++\chi &=-\tilde{G}_m^++\tilde{\Theta}\ .
\label{massivegchiid}
\end{align}
This relation can be derived by using the  following relations
\begin{subequations}
\begin{align}
\chi &=\frac{-i (1-\mu^2)}{2\pi }\int_{\tilde{\mathcal{\zeta}}_c}^{\tilde{\mathcal{\zeta}}}d\tilde{\mathcal{\zeta}}'\frac{\tilde{\mathcal{\zeta}}'^{\frac{1}{2}}}{(\tilde{\mathcal{\zeta}}'-\mu^2)(\tilde{\mathcal{\zeta}}'-1)}\\
\bar{\Theta}&=\frac{ (1-\mu^2)}{2\pi i}\int_{\tilde{\mathcal{\zeta}}_c}^{\tilde{\mathcal{\zeta}}}d\tilde{\mathcal{\zeta}}'\frac{\tilde{\mathcal{\zeta}}'^{-\frac{1}{2}}}{(\tilde{\mathcal{\zeta}}'-\mu^2)(\tilde{\mathcal{\zeta}}'-1)}\\
\tilde{G}_{\omega}^+&=\frac{i (1-\mu^2)}{2\pi}\int_{\tilde{\mathcal{\zeta}}_c}^{\tilde{\mathcal{\zeta}}}d\tilde{\mathcal{\zeta}}'\left\{\frac{\tilde{\mathcal{\zeta}}'^{\frac{1}{2}}}{(\tilde{\mathcal{\zeta}}'-\mu^2)(\tilde{\mathcal{\zeta}}'-1)}-\frac{(G_0^+[1])^2}{(\tilde{\mathcal{\zeta}}'-\mu^2)(\tilde{\mathcal{\zeta}}'-1)(G_0^+[\tilde{\mathcal{\zeta}}'])^2}\right\}\\
\tilde{G}_{m}^+ &=\frac{ (1-\mu^2)}{2\pi i}\int_{\tilde{\mathcal{\zeta}}_c}^{\tilde{\mathcal{\zeta}}}d\tilde{\mathcal{\zeta}}'\left\{\frac{\tilde{\mathcal{\zeta}}'^{-\frac{1}{2}}}{(\tilde{\mathcal{\zeta}}'-\mu^2)(\tilde{\mathcal{\zeta}}'-1)}-\frac{(G_0^+[1])^2}{(\tilde{\mathcal{\zeta}}'-\mu^2)(\tilde{\mathcal{\zeta}}'-1)(G_0^+[\tilde{\mathcal{\zeta}}'])^2}\right\}\ .
\end{align}
\label{chitheta}
\end{subequations}
Let us define $x=\frac{2\tilde{\mathcal{\zeta}}-1-\mu^2}{1-\mu^2}$. Then one can write the sum in the LHS or RHS of \eqref{massivegchiid} as follows
\begin{align}
\tilde{G}_{\omega}^++\chi &=\frac{1}{i\pi}\int_{x_c}^x\frac{1}{(x^2-1)P_{-\nu}(x)^2}\nonumber\\
&=\frac{i}{\pi}\frac{Q_{-\nu}(x)}{P_{-\nu}(x)}+i\cot(\pi\nu)\ 
\label{sumgchi}.
\end{align}
Using the derivative expanded solution for the massive case, let us analyse the radial integrals as given in \eqref{inf2}. For this, first let us note the asymptotic behaviour of $G_0^+$ at large $\tilde{\zeta}$
\begin{align}
P_{-\nu}\left(\frac{2\tilde{\zeta}-1-\mu^2}{1-\mu^2}\right)&\xrightarrow[\tilde{\zeta}\rightarrow \infty]{} \frac{\tilde{\zeta}^{\nu-1}}{1-\mu^2}\
\label{asympG0} .
\end{align}
Using \eqref{asympG0}, we obtain the following asymptotic expression for the radial integrals. For odd $k=2l+1$ from equation \eqref{inf2}, we get
\begin{align}
\mathcal{F}^b_{n,2l+1}\sim \frac{1}{4^k}\int^{r_c} dr\: r\: (G_0^+)^n\sim\frac{1}{4^k}r_+^2\left(\frac{\tilde{\zeta}_c}{1-\mu^2}\right)^{(\nu-1)n}\frac{\tilde{\zeta}_c}{2(\nu-1)n+2}\ .
\end{align}
Hence this is divergent for $\nu>1-\frac{1}{n}$. For even $k=2l$, using \eqref{asympG0} and \eqref{sumgchi}, we get the following divergence
\begin{align}
\mathcal{F}^b_{n,2k}\sim \frac{2k}{4^{k-1}}\int^{r_c}dr\: r\: (G_0^+)^n\left(\chi+\tilde{G}_{\omega}^+\right)\sim \frac{2k}{4^{k-1}}\frac{1}{(1-\mu^2)^{(\nu-1)n-2\nu+1}}\frac{r_+^2}{i\pi}\frac{\tilde{\zeta}_c^{(\nu-1)(n-2)}}{2(\nu-1)(n-2)(1-2\nu)}
\label{div1}.
\end{align}
This integral is thus divergent when $\nu>1$ for $n\geq3$. However, for $\nu<1$ that corresponds to relevant operator in the dual CFT, there is no divergence for even $k$.
\addcontentsline{toc}{section}{References}
\bibliography{v1} 
\bibliographystyle{JHEP}
\end{document}